\documentclass[apj,numberedappendix]{emulateapj}
\usepackage{epstopdf}
\usepackage{wrapfig}
\usepackage{natbib}
\bibpunct{(}{)}{;}{a}{}{,}

\usepackage{color}
\definecolor{darkgreen}{RGB}{0,142,128}
\definecolor{darkblue}{RGB}{0,100,170}

\usepackage{hyperref}
\hypersetup{colorlinks,citecolor=blue,linkcolor=blue}

\begin{document}
\title{Global solar magnetic field organization in the outer corona: influence on the solar wind speed and mass flux over the cycle.}

\author{Victor R\'eville$^*$}
\thanks{* New affiliation: UCLA Earth, Planetary and Spaces Sciences, 595 Charles Young Drive East, Los Angeles CA 90035; vreville@epss.ucla.edu}
\author{Allan Sacha Brun}
\affil{Laboratoire AIM, DSM/IRFU/SAp, CEA Saclay, 91191 Gif-sur-Yvette Cedex, France; victor.reville@cea.fr}

\begin{abstract}
The dynamics of the solar wind depends intrinsically on the structure of the global solar magnetic field, which undergoes fundamental changes over the 11-yr solar cycle. For instance, the wind terminal velocity is thought to be anti-correlated with the expansion factor, a measure of how the magnetic field varies with height in the solar corona, usually computed at a fixed height  ($\approx 2.5 R_{\odot}$, the source surface radius which approximates the distance at which all magnetic field lines become open). However, the magnetic field expansion affects the solar wind in a more detailed way, its influence on the solar wind properties remaining significant well beyond the source surface. We demonstrate this using 3D global MHD simulations of the solar corona, constrained by surface magnetograms over half a solar cycle (1989-2001). A self-consistent expansion beyond the solar wind critical point (even up to $10R_{\odot}$) makes our model comply with observed characteristics of the solar wind, namely, that the radial magnetic field intensity becomes latitude independent at some distance from the Sun, and that the mass flux is mostly independent of the terminal wind speed. We also show that near activity minimum, the expansion in the higher corona has more influence on the wind speed than the expansion below $2.5 R_{\odot}$.
\end{abstract}


\section{Introduction} 
\label{intro}

Relating in-situ measurements of the solar wind and remote observations of the Sun and the solar corona has been a continuous effort for the past 60 years, starting with the pioneering work of \citet{Parker1958}. Within this standard wind model, the observed $\sim 1$ MK coronal temperature is able to dynamically expand an hydrodynamical solar atmosphere to supersonic speed around $400$ km/s at one astronomical unit (AU). The solar magnetic field has further been found to play a fundamental influence on the structure of the corona and the solar speed. The solar wind speed, density, magnetic field and currents see changes following the solar sunspot cycle of 11 years \citep[][]{Neugebauer1975}. Through this cycle, the solar magnetic field observed at the photosphere changes its geometry and its amplitude with the minimum of activity showing a dipolar configuration and the maximum of activity a quadrupolar or multipolar configuration \citep{DeRosa2012}.

With its out of the ecliptic orbit, the \textit{Ulysses} spacecraft has revealed, over a full cycle, the 3D structure of the solar wind and its relation to the magnetic field geometry \citep{McComas1998,McComas2003}. The two components of the solar wind, the slow (around $400$ km/s at $1$ AU) and the fast (around $800$ km/s at $1$ AU) are believed to originate nearby and away from closed coronal loops respectively. In-situ measurements have revealed that the mass flux in the solar wind at 1 AU is between $2 \times 10^8$ cm$^{-2}$ s$^{-1}$ in the fast streams and $3-4 \times 10^8$ cm$^{-2}$ in slow streams \citep[see e.g.][for observations made with \textit{Helios} and \textit{Ulysses} respectively]{Schwenn1983,Goldstein1996}. Modeling the repartition of closed loops and open regions in the corona is thus crucial to predict the solar wind characteristics at Earth orbit. \textit{Ulysses} observations have also revealed that the radial magnetic field is constant with latitude \citep{SmithBalogh1995,Smith2011}. This empirically validates that magnetic field measurements made in the ecliptic plane are a good proxy for the total flux emerging from the Sun. 

The well known potential field source surface model \citep[PFSS, see][]{AltschulerNewkirk1969,Schatten1969,SchrijverDeRosa2003} gives, in an extremely computationally efficient way, a structure for the coronal magnetic field from observed magnetograms. Assuming a current free corona, up to the source surface radius where the field becomes radial, the PFSS model yields the location and the area of coronal holes as well as the open regions polarity (and thus the location of the heliospheric current sheet). The source surface radius is usually set around $2.5 R_{\odot}$, even though the optimal value is likely to vary over the solar cycle, in order to match, for instance, the variation of the magnetic open flux \citep{Lee2011,Arden2014}. 

The PFSS model yields the value of the expansion factor $f_{\mathrm{exp}}$, \textit{i.e.} how a given flux tube expansion deviates from a pure radial field, up to the source surface. This parameter has been observed to be anti-correlated with the terminal wind speed \citep{WangSheeley1990}. The PFSS model has thus been used, combined with other models able to propagate the solution to 1 AU, to predict the solar wind speed and IMF polarity from magnetograms, notably within the Wang-Sheeley-Arge (WSA) model \citep{ArgePizzo2000,Arge2003}.  The relation obtained between the expansion factor at the source surface and the terminal wind speed in the WSA model needs, however, to be calibrated and corrected with observations. The PFSS model indeed fails to reproduce the interaction between the magnetic field and flow dynamics, yielding for instance a latitude dependent profile of the radial magnetic field.

Full magnetohydrodynamics (MHD) models are consequently necessary to obtain a more coherent picture. \citet{Usmanov1993} first used observed magnetograms within a full MHD model to analyze the properties of the solar corona and wind. \citep{Mikic1999} have shown that MHD models constrained by photospheric magnetograms yield a good description of the structure, shape and size of coronal loops as they appear during eclipses. The radial field is also observed to be independent of latitude in these models, as a consequence of the latitudinal balance between the thermal and dynamic pressure of the wind and the magnetic forces, realized at $\beta \geq 1$, in the higher corona \citep[see][for a detailed discussion]{Usmanov2000}. This effect has an influence on the expansion of the field lines beyond the canonical value of the source surface, and thus cannot be captured by the PFSS. \citet{Pinto2016} studied, in an axisymmetric setup, the expansion of the field lines over the solar cycle and showed that the link between the expansion in the whole domain and the wind terminal speed clearly depends on the cycle phase.
 
The purpose of this paper is to study the 3D structure of the solar corona and the heliospheric magnetic fields at a few solar radii from the Sun, over the solar cycle, and to yield a new understanding on the structure of flux tube that is obtained with MHD models. Although many papers have already been dedicated to this topic, they focus mainly on the structure of closed loops or open regions close to the Sun. For instance, \citet{Riley2006,Reville2015b} have shown that below the source surface, the agreement between MHD simulations and the PFSS were acceptable, both providing similar values of the open magnetic flux \citep[if the value of the source surface is well chosen, see][]{Reville2015b}. Deviations beyond the source surface are, however, significant and have consequences on the solar wind speed predictions as shown recently by \citet{Cohen2015SoPh}. The geometry of the field lines evolve continuously from the surface to and within the interplanetary medium. It is misleading to think that they become strictly radial near the Sun, whether it is beyond a source surface radius around $2.5 R_{\odot}$ or even beyond the sonic or the Alfv\'en point, located further on. We demonstrate this performing MHD simulations using as an input WSO magnetograms, one per year between 1989 and 2001. We observe a reorganization of the field that is taking place between the surface and about 10 solar radii, \textit{i.e.} far beyond the typical value of the source surface used for the Sun.

In Section \ref{sec:num} we detail our numerical model, and then describe the 13 simulations we have made using Wilcox magnetograms to cover half a 22-yr solar cycle for the period 1989-2001. The global structure of the corona is described in Section \ref{sec:glob}, and the average location of what we define as an optimal source surface radius, the sonic point and the Alfv\'en point are analyzed over the cycle. Section \ref{sec:pfssmhd} shows that the latitude independent $B_r$ established in our simulations is responsible for an enhanced field line expansion beyond the optimal source surface radius. In Section \ref{sec:wsc} we look at the influence of the expansion on the wind speed, and find that the well known anti-correlation between the total expansion factor and wind speed is reversed between the source surface and the higher corona during solar minimum. We also show the evolution of the mass flux with height and underline the importance of the expansion both below and beyond the source surface to match observations. We discuss these results and set the limitations of our model in Section \ref{sec:disc}.

\section{Numerical setup}
\label{sec:num}

\subsection{MHD model}
The numerical simulations reported here are done using the PLUTO code \citep{Mignone2007}. We follow the exact same procedure as in \citet{Reville2016} and solve the time dependent ideal (non-resistive) magnetohydrodynamics (MHD) equations: 
\begin{equation}
\label{MHD_1}
\frac{\partial}{\partial t} \rho + \nabla \cdot \rho \mathbf{v} = 0,
\end{equation}
\begin{equation}
\label{MHD_2}
\frac{\partial}{\partial t} \rho \mathbf{v} + \nabla \cdot (\rho \mathbf{vv}-\mathbf{BB}+\mathbf{I}p) = - \rho \nabla \Phi + \rho \mathbf{a},
\end{equation}
\begin{equation}
\label{MHD_3}
\frac{\partial}{\partial t} (E + \rho \Phi)  + \nabla \cdot ((E+p+\rho \Phi)\mathbf{v}-\mathbf{B}(\mathbf{v} \cdot \mathbf{B})) = \rho \mathbf{v} \cdot \mathbf{a},
\end{equation}
\begin{equation}
\label{MHD_4}
\frac{\partial}{\partial t} \mathbf{B} + \nabla \cdot (\mathbf{vB}-\mathbf{Bv})=0,
\end{equation}
where $E \equiv  \rho \epsilon + \rho v^2/2 + B^2/2$ is the total energy, $\mathbf{B}$ is the magnetic field, $\rho$ is the mass density, $\mathbf{v}$ is the velocity field, $p = p_{\mathrm{th}} +B^2/2$ is the total (thermal plus magnetic) pressure and $\mathbf{I}$ is the identity matrix. 

The potential $\Phi$ accounts for the gravitational attraction of the star and $\mathbf{a}$ is a source term which contains the Coriolis and centrifugal forces as we solve the equations in a rotating frame at $\Omega = 2.6 \times 10^{-6}$ rad/s (the 28 days average rotational period of the Sun). The numerical scheme is a Harten, Lax, van Leer and Einfeldt \citep[HLLE, see][]{Einfeldt1988} Riemann solver, combined with a second order linear reconstruction method and minmod slope limiter. $\nabla \cdot \mathbf{B} = 0$ is ensured by a constraint transport method \citep{EvansHawley1988,BalsaraSpicer1999}.

The closure equation for the system relates the internal energy $\epsilon$ to the thermal pressure and density $(p,\rho)$ as follows:
\begin{equation}
\epsilon = \frac{p}{\rho(\gamma-1)},
\end{equation}
with $\gamma=1.05$, the ratio of specific heats, which differs from the usual value of $5/3$ for a hydrogen gas in order to mimic the extended coronal heating. Polytropic models use mostly three input parameters to describe the thermodynamics of the outflow. This value of $\gamma$ has been widely used in the literature, and its combination with a base coronal temperature of $T_{\odot} = 1.5 \times 10^6$ K has shown to reproduce accurately the structure of closed loops as observed during eclipses \citep{Mikic1999} or with a coronagraph.  The base density is set for all simulations at $n_{\odot} = 10^8$ cm$^{-3}$ in order to get a total integrated mass flux around the upper canonical value of  $3 \times 10^{-14} M_{\odot}$/yr. Although these input values could be a function of the period of the solar cycle, with for instance higher temperature near solar maximum, we chose to keep them constant for all our simulations for simplicity. More advanced heating processes, which include the propagation and dissipation of Alfv\'en waves \citep{Usmanov2000,Lionello2009,Sokolov2013,vanDerHolst2014}, could also result in different effective heating for different regions at the same epoch. However, we believe that this does not change the global outcome of the reflexion put forward in this paper (see Section \ref{sec:disc}).

The domain is a 3D cartesian box of $60 R_{\odot}$ cubed, with $448$ cells in each direction. The grid is refined in the domain $[-1.5 R_{\odot}, 1.5 R_{\odot}]^3$, and stretched beyond. The solar interior is used as a boundary condition (identified as the coronal base), where a specific procedure is used to conserve MHD invariants in the domain \citep[see][for a more detailed discussion on the boundary conditions]{Reville2016}. In this region the surface magnetic field is imposed from observations, and reconstructed with a potential field source surface model (PFSS) at $t=0$. 

\subsection{WSO Magnetograms}

\begin{figure}
\center
\includegraphics[width=3in]{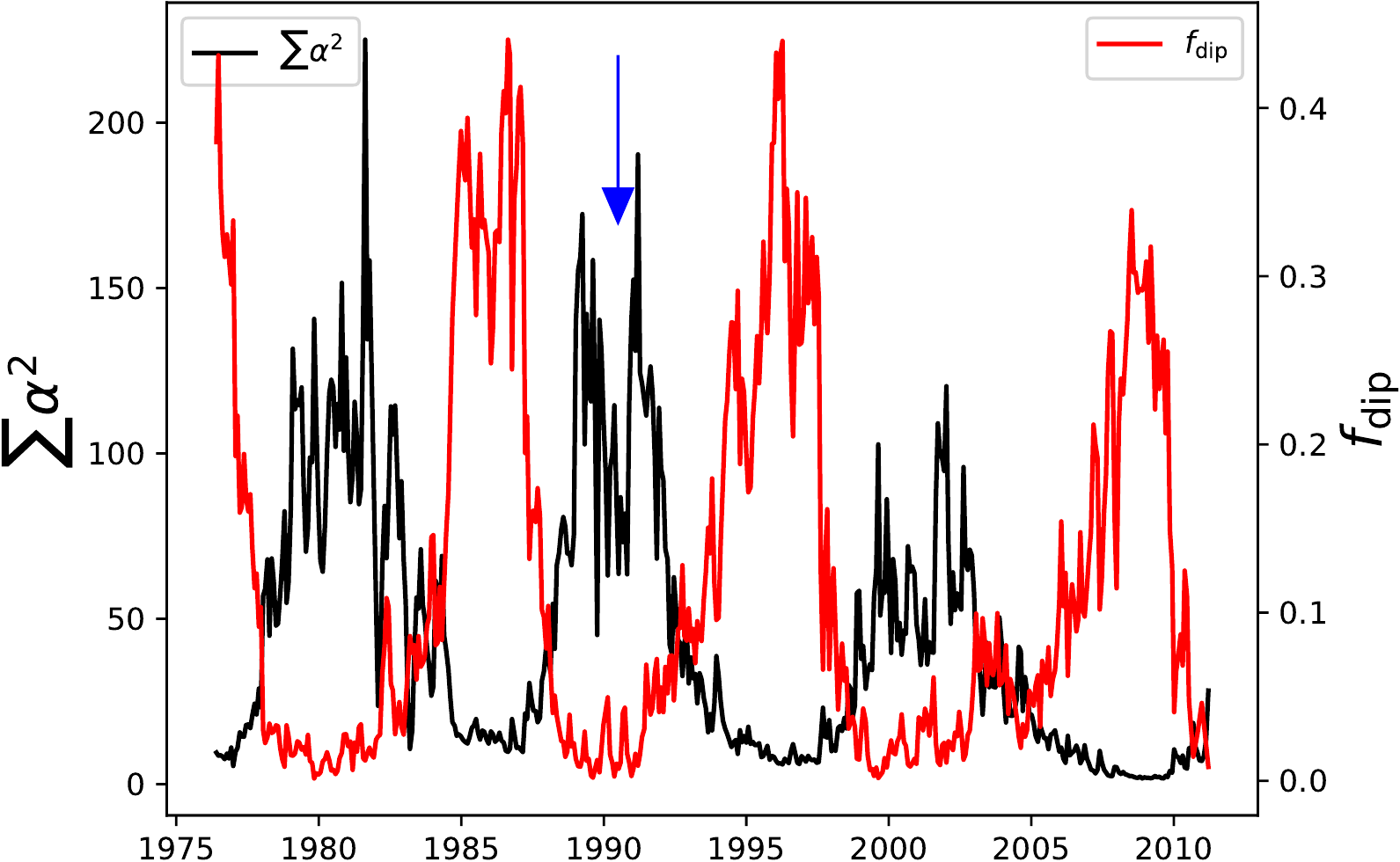}
\caption{Comparison between the surface magnetic energy (in black) of the WSO maps and the dipole component ($f_{\mathrm{dip}}$ in red) evolution for the whole WSO activity period.}
\label{scmap}
\end{figure}

Our study is based on Wilcox Solar Observatory (WSO) magnetograms \citep{Scherrer1977}. These synoptic magnetograms identify the magnetic field in the line of sight with the radial component of the surface field, and we take from the work of \citet{DeRosa2012} the spherical harmonics decomposition of $B_r$ for each Carrington rotation:

\begin{equation}
B_r (1 R_{\odot}, \theta, \varphi) = \sum_{\ell,m} \alpha_{\ell,m} Y_{\ell,m} (\theta,\varphi),
\end{equation}
where the $Y_{\ell,m}$ are the scalar spherical harmonics functions. 
The Wilcox solar observatory has measured continuously the solar magnetic field since 1976 (CR 1642). Figure \ref{scmap}, shows the typical properties of the solar magnetic field for the past $40$ years. The magnetic energy at the surface $\sum \alpha_{\ell,m}^2$ oscillates, with a period of $\approx 11$ years, half the full magnetic cycle that includes the reversal. It reaches three maximums around 1980, 1991, and 2003, with a lower amplitude for the last cycle, which was also observed counting sunspots at the surface.

In Figure \ref{scmap}, we also show the parameter
\begin{equation}
f_{\mathrm{dip}} = \frac{\alpha_{1,0}^2+\alpha_{1,1}^2}{\sum_{\ell,m} \alpha_{\ell,m}^2},
\end{equation}
which quantifies the energy of the dipole component (axisymmetric and non axisymmetric) over the total energy of the surface radial field in its spherical harmonics decomposition. Figure \ref{scmap} illustrates the anti-correlation between the total magnetic field energy and the energy contained in the dipolar components over the whole period. Equivalently, we say that the Sun is almost dipolar at minimum of activity, and multipolar near maximum of activity \citep[see][for a deeper analysis on the dynamo families along the solar cycle]{DeRosa2012}.

\begin{deluxetable}{c|c|c|c|c}
  \tablecaption{WSO magnetograms}
  \tablecolumns{5}
  \tabletypesize{\scriptsize}
  \tablehead{
    \colhead{CR} &
    \colhead{year} &
    \colhead{$r_{\mathrm{ss},\mathrm{opt}}$} &
    \colhead{$\langle r_c \rangle$} &
    \colhead{$\langle r_A \rangle$}
  }

  \startdata
  1811 & 1989.02 & 2.37 & 5.53 & 5.62 \\
  1824 & 1989.99 & 2.67 & 5.71 & 6.60 \\
  1837 & 1990.97 & 2.30 & 5.42 & 5.47 \\
  1850 & 1991.94 & 2.90 & 5.82 & 7.08 \\
  1863 & 1992.91 & 2.82 & 5.67 & 6.54 \\ 
  1876 & 1993.88 & 2.52 & 5.52 & 6.11 \\
  1889 & 1994.85 & 2.52 & 5.57 & 6.28 \\
  1902 & 1995.82 & 2.30 & 5.46 & 5.98 \\
  1915 & 1996.79 & 2.30 & 5.53 & 6.07 \\
  1928 & 1997.76 & 2.22 & 5.53 & 5.86 \\
  1941 & 1998.73 & 2.00 & 5.44 & 5.50 \\
  1954 & 1999.70 & 2.22 & 5.32 & 5.40 \\
  1967 & 2000.67 & 2.08 & 5.40 & 5.40 \\
  \enddata
  \tablecomments{Carrington rotation, decimal year of the WSO magnetograms used in our 13 simulations. The optimal source surface as well as the average spherical sonic point and Alfv\'en point are given for each case.}
  \label{table1}
\end{deluxetable}

We chose to perform a set of 13 simulations, focusing on the period 1989-2001, to cover half a 22-yr cycle during the beginning of the \textit{Ulysses} spacecraft observations. We reconstruct the initial solution with a source surface radius of $15 R_{\odot}$, using the coefficients for the spherical harmonics up to $\ell=15$ (\textit{i.e.} $135$ coefficients). We also performed some of the cases with WSO magnetograms' maximum resolution ($\ell=60$), and did not see any significant difference for the results presented in this paper. The correction factor that sometimes multiplies the WSO magnetograms is not applied here \citep[see][]{Svalgaard1978}. Table \ref{table1} lists the Carrington rotations associated with the Wilcox magnetograms used in our 13 simulations. The corresponding decimal year is shown and the period considered starts around maximum of activity in 1989, goes through a minimum around 1995-1996 and goes back to a maximum state of activity late in year 2000.

\section{Global structure of the corona}
\label{sec:glob}

\begin{figure*}
\begin{center}
\begin{tabular}{ccc}
CR 1824 & CR 1850 & CR 1876\\ 
\includegraphics[scale=0.15]{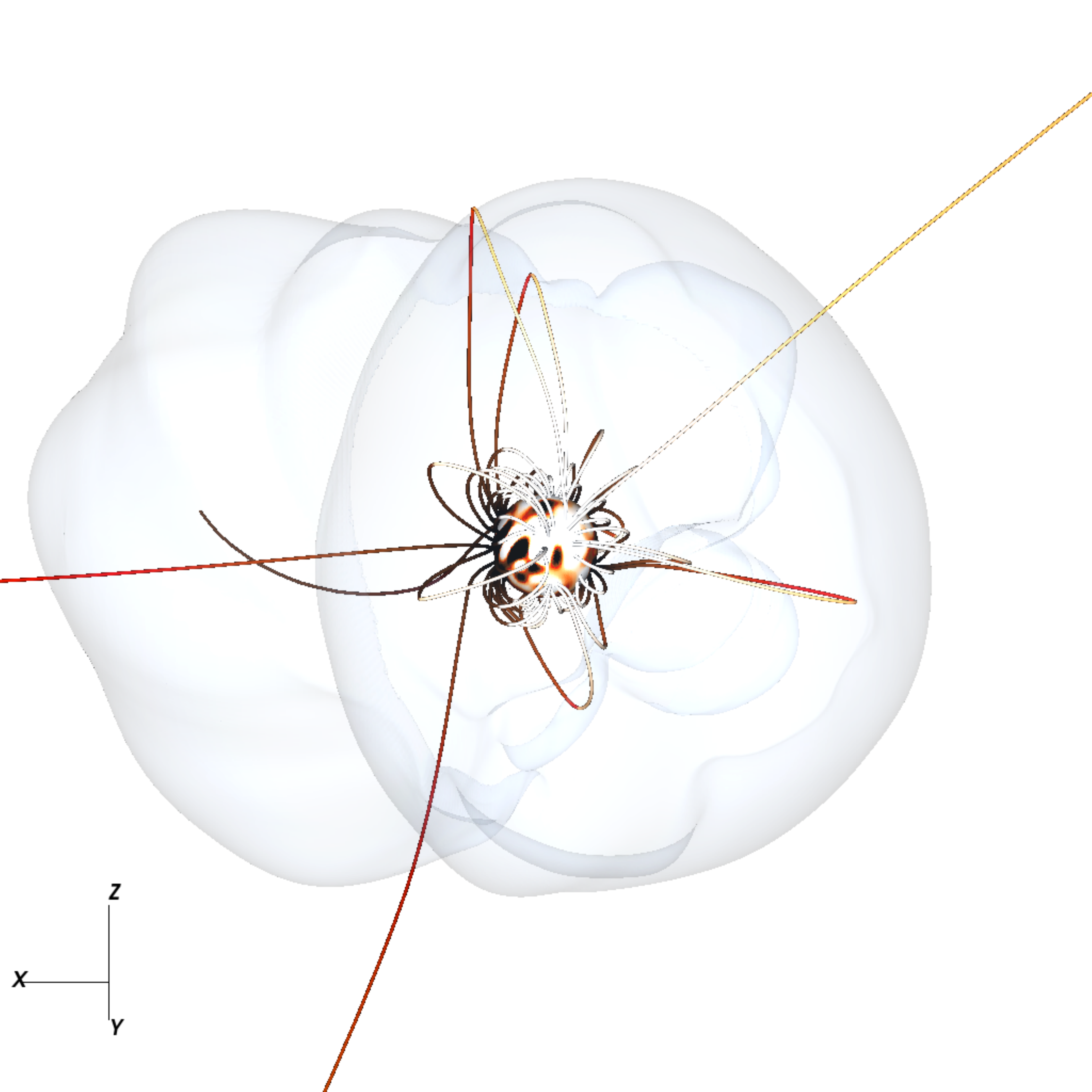} & \includegraphics[scale=0.15]{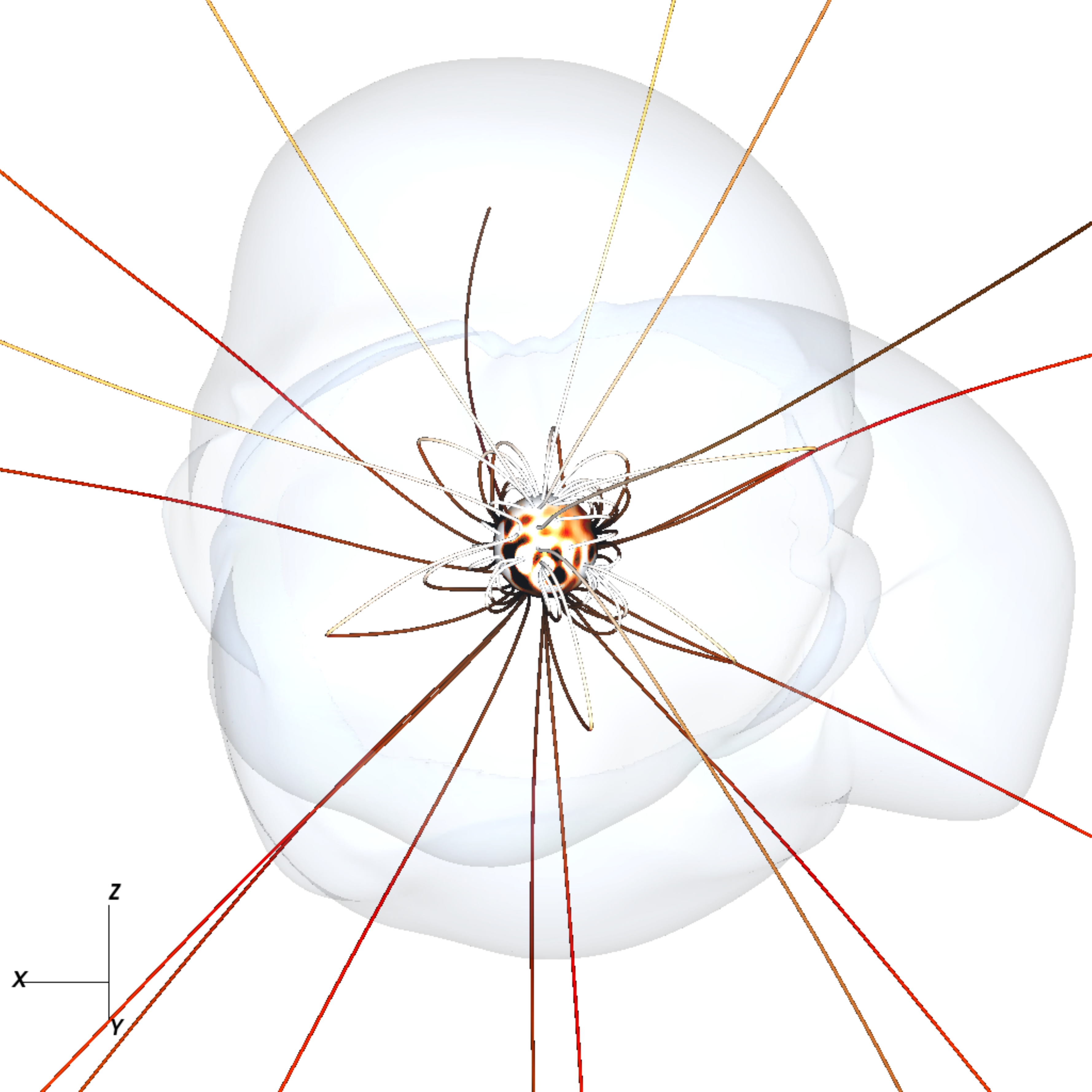} & \includegraphics[scale=0.15]{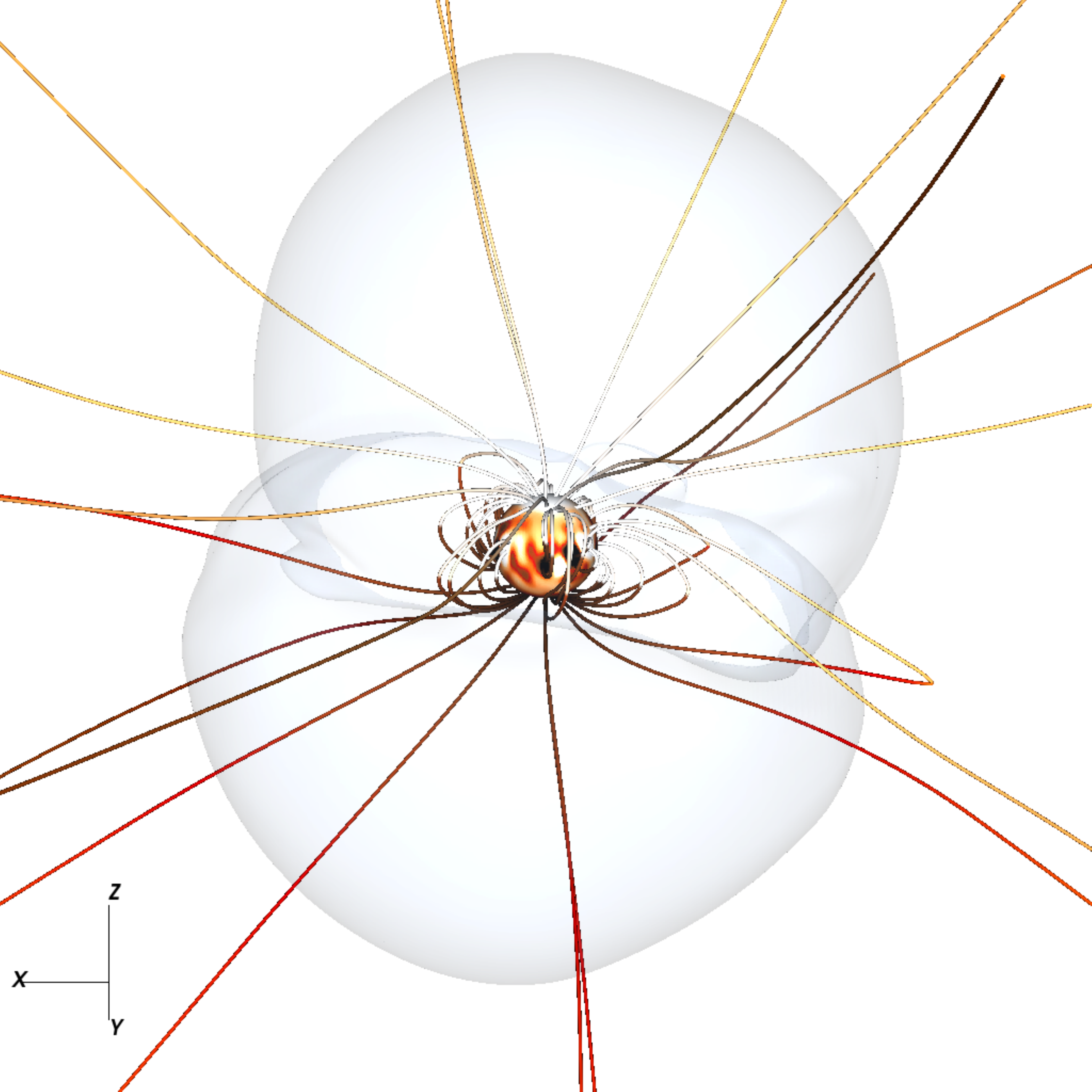} \\
CR 1902 & CR 1928 & CR 1954\\
 \includegraphics[scale=0.15]{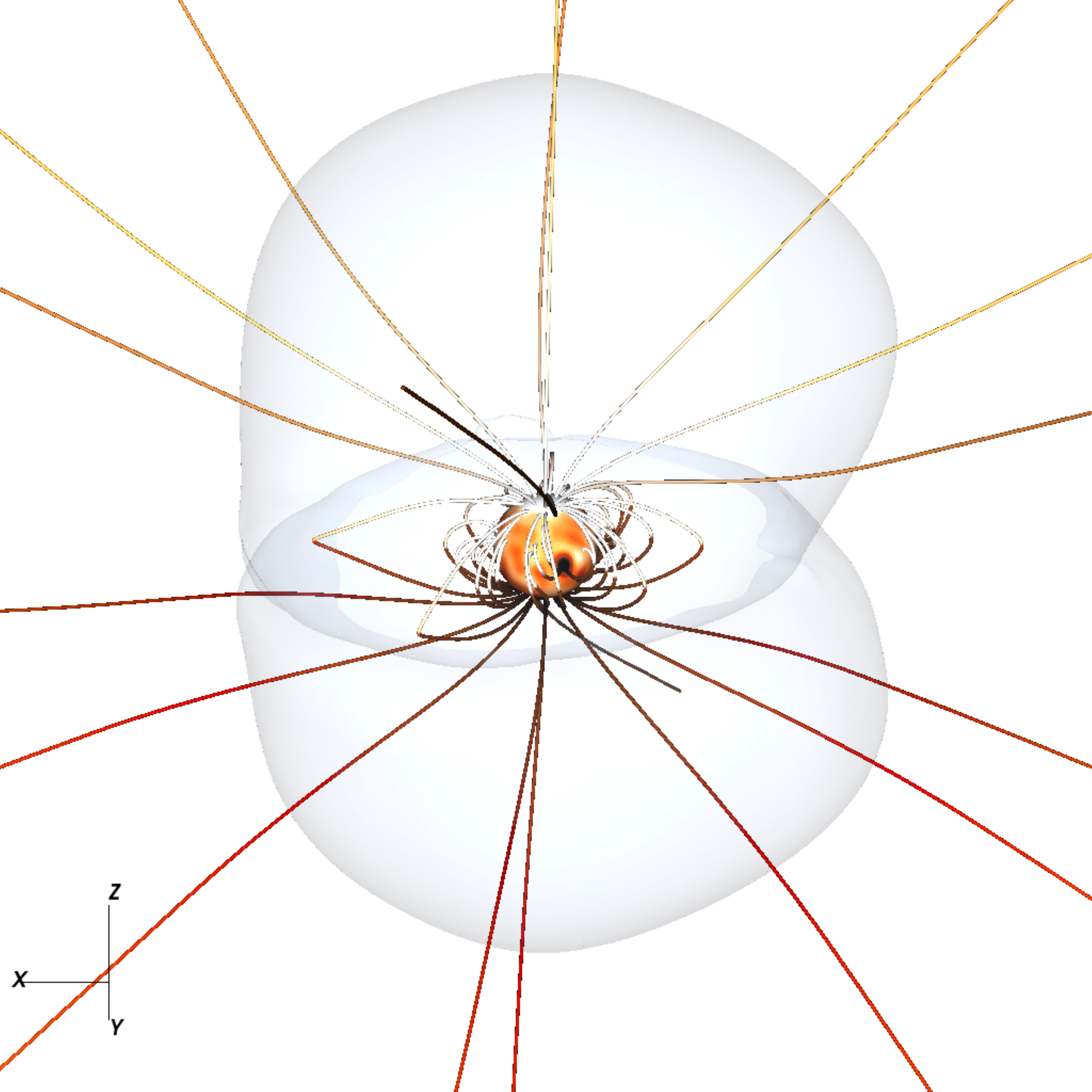} & \includegraphics[scale=0.15]{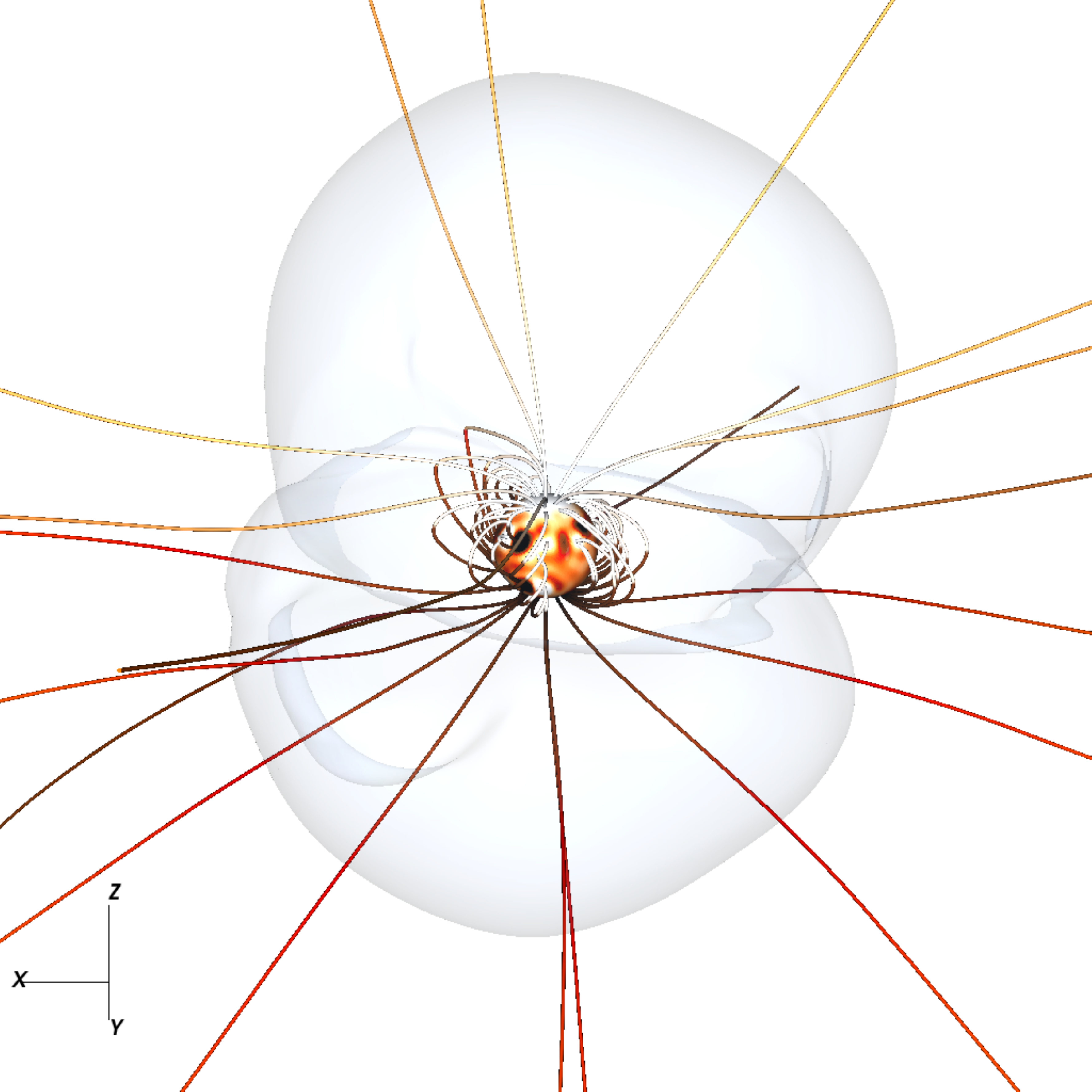} & \includegraphics[scale=0.15]{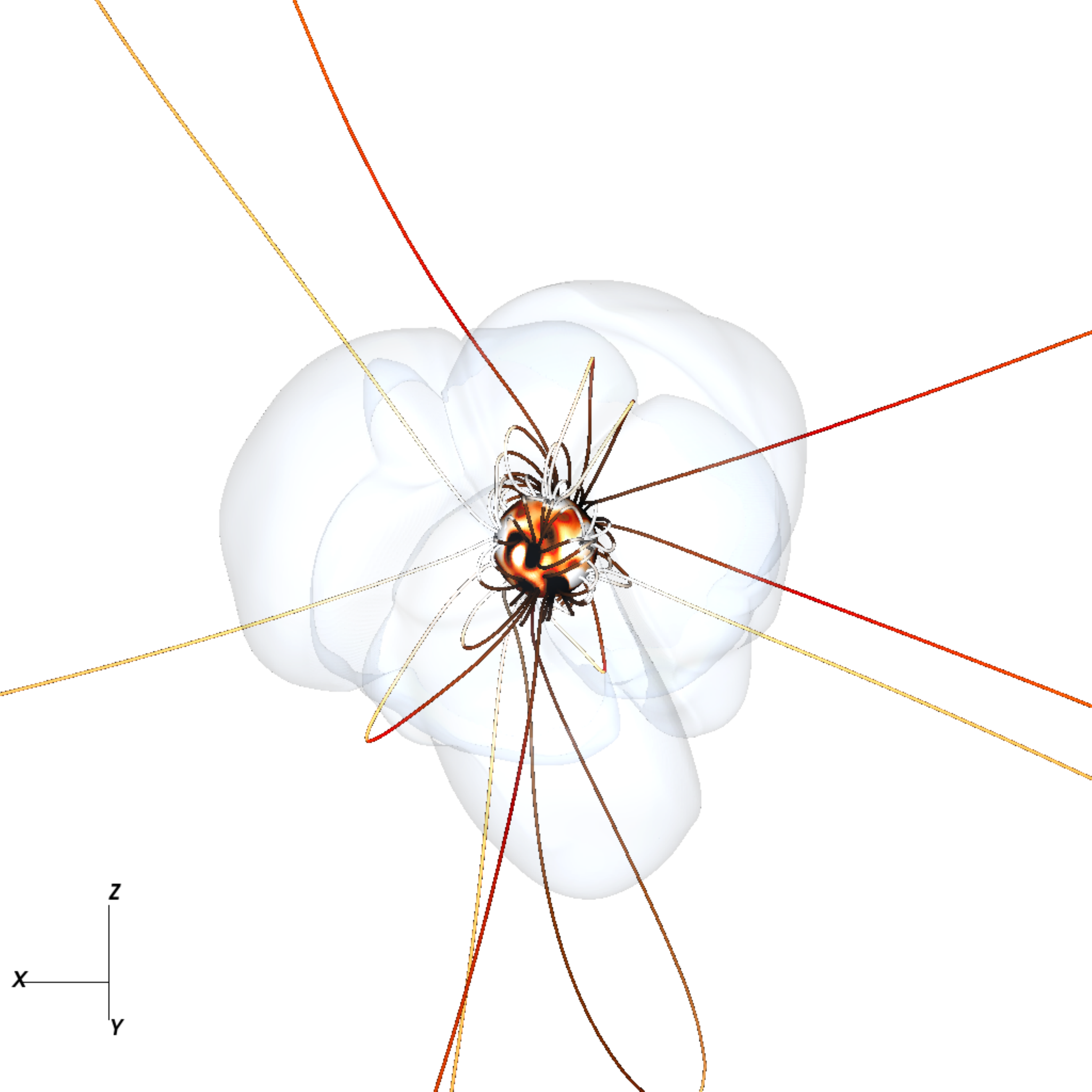} \\
\end{tabular}
\end{center}
\caption{Shown are 3D rendering of 6 of our MHD solutions of the solar wind using WSO maps. The field lines are shown red or yellow depending on the polarity of the field. The Alfv\'en surface is shown in translucent grey. The surface radial magnetic field is shown using color contour maps. CR 1824 and CR 1954 represent clear cases of solar maximum with a non-axisymmetric dipole and a complex multipole structure respectively.}
\label{AlfSurf}
\end{figure*}

The results of the simulations are shown in Figure \ref{AlfSurf}, for six simulations out of our 13 cases.  The steady state is characterized by a balance between the outward thermal pressure, which creates the solar wind, and the magnetic forces that can contain the plasma in co-rotation with the Sun within closed loops. Using observed magnetograms all over the cycle creates a time varying and complex distribution of loops and open regions. 

Around minimum of activity, the structure is mostly dipolar, and the Alfv\'en surface -where the solar wind reaches the Alfv\'en speed $v_A=B/\sqrt{\mu_0 \rho}$- shown in grey shades, has its usual bilobal structure. Each lobe corresponds to the northern and southern hemispheres coronal holes. The axisymmetric dipolar component is, in effect, the dominant mode during those epochs. Near maximum of activity (\textit{i.e.} for CR 1824 and CR 1954) we observe a more complex structure with smaller lobes in the Alfv\'en surface that corresponds to coronal holes and polarity connected open regions. The Alfv\'en point always touches the cusp of the helmet streamers, and is thus the closest to the star near the current sheets. In Figure \ref{AlfSurf} the polarity of the field lines is shown reddish for positive, yellowish for negative. When the magnetic field geometry is mostly dipolar, we observe a clear separation between positive and negative polarities -separation embodied by the heliospheric current sheet-, located in the equatorial plane near the activity minimum. This clear separation disappears around maximum where current sheets can be located at all latitudes. 

\begin{figure}
\center
\includegraphics[width=3in]{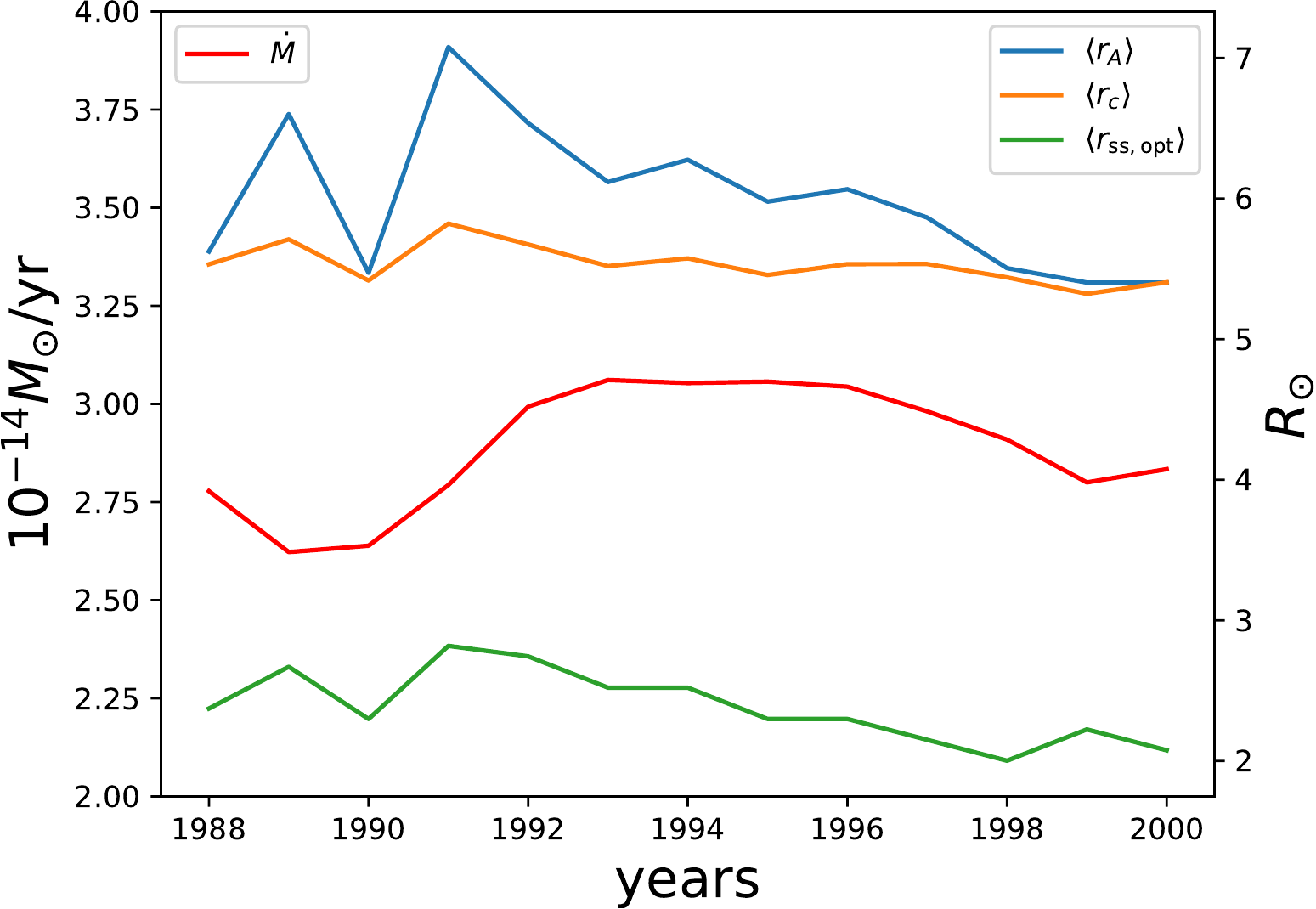}
\caption{Evolution of the mass loss rate in the simulations (in red, left y-axis) and of the optimal source surface radius, the average spherical sonic point, and the average spherical Alfv\'en point over the cycle (right y-axis).}
\label{radii}
\end{figure}

In Figure \ref{radii} we show the evolution of the total mass loss rate $\dot{M}= \int \rho \mathbf{v} \cdot d\mathbf{S}$. The mass loss rate varies by some $20\%$, between $2.6 \times 10^{-14} M_{\odot}$/yr and $3.1 \times 10^{-14} M_{\odot}$/yr, within the observational constraints. The mass loss minimum is located around year 1990. The maximum is reached in year 1993, between solar activity maximum and solar activity minimum, then slowly decreases until 1999 and raises again in 2000. The correlation between the mass loss variations and the structure of the solar magnetic field is subtle, but $\dot{M}$ is globally anti-correlated with the magnetic energy (see Figure \ref{scmap}). Its variations match the observations of the \textit{Ulysses} spacecraft as shown in \citet{McComas2003,McComas2008}. The \textit{in-situ} data starts with a minimum value of $\dot{M}$ around $2.3 \times 10^{-14} M_{\odot}$/yr in 1991 and reaches a maximum of $3.0 \times 10^{-14} M_{\odot}$/yr in 1992-1993. The mass loss then slowly decreases until the new solar activity maximum phase in year 2000, where it starts increasing again. We computed these values from the dynamical pressure $\rho v^2$ and wind speed given in \citet{McComas2008}, assuming a spherical symmetry.

Given the topology changes of the solar surface magnetic field observed over the cycle phase we expect the variation of the overall size of the Alfv\'en surface to be correlated with it. In \citet{Reville2015a} we showed that the location of the Alfv\'en was strongly influenced by the topology of the surface magnetic field, and that it was a decreasing function of the complexity -the degree $\ell$ of the component in the spherical harmonics decomposition of the field \citep[see also][]{Finley2017}. Previous studies such as \citet{PintoBrun2011} were predicting a factor 4 between minimum and maximum of activity, due to the strong decreases of the dipole component at maximum. The strong modulation of $f_{\mathrm{dip}}$ (more than a factor 10) would therefore suggest a significant modulation of the Alfv\'en radius. However, as shown in Table \ref{table1}, and in Figure \ref{radii}, this variation is at most of $24\%$ in average (the variation measured for all $(\theta, \phi)$ is larger). The clear anti-correlation between $f_{\mathrm{dip}}$ and the magnetic energy shown in Figure \ref{scmap} is certainly the reason for this small variation and both the (increase) decrease of the dipole and the (decrease) increase of the magnetic energy must be looked at to understand the variation we obtain.

For instance, the maximum extension of the Alfv\'en surface is not exactly located at the minimum of activity but rather between maximum and minimum for CR 1850 when the dipole component is regaining strength, and the surface field has not yet fallen into its lowest energy levels. Similarly, the smallest average Alfv\'en surface are reached for CR 1954 and CR 1967, two years before the maximum of activity, when the dipole component and the magnetic energy are both weak. Also, the sharp drop observed for CR 1837 is due to the drop in amplitude around year 1991, shown in Figure \ref{scmap} with a blue arrow.

Hence, taking into account the right variation of the field's energy, or amplitude, is crucial to correctly capture the variation of the Alfv\'en surface over the cycle. This is likely the main difference between these results and the results of \citet{PintoBrun2011}, where the surface magnetic field was taken from a kinematic dynamo code, and was probably not recovering the right amplitude variation. They also find the mass loss to be maximum during maximum of activity. The integrated mass loss is strongly influenced by the spatial distribution of slow, dense wind and the value at maximum of activity is probably enhanced by the axisymmetry of their model. This feature has been advanced in other works such as \citet{Wang1998}, which based its estimates on PFSS extrapolations and thus might not be fully consistent at estimating the mass flux (see Section \ref{sec:wsc}). Nevertheless, the wind driving is likely to be an increasing function of the surface magnetic field energy \citep[through the Alfv\'en waves energy flux for instance, see e.g.][]{Suzuki2006,Cranmer2007,Pinto2017}. This would certainly amplify the variation of the Alfv\'en surface over the cycle, with a clearer peak near activity minimum, and we can consider our results, obtained with fixed acceleration processes for the wind, as a minimum case.

In addition to the average Alfv\'en radius, we show in Figure \ref{radii}, the evolution of the average critical point, where the wind becomes supersonic, and of the optimal source surface radius $r_{\mathrm{ss,opt}}$. The latter is defined from our previous work \citep{Reville2015b}, as the source surface radius that makes a PFSS model match the open flux of the simulation. We see that $r_{\mathrm{ss,opt}}$ is always smaller that the Alfv\'en radius and it evolution follows the same variation than the average Alfv\'en radius. The peak in both quantities is for CR 1850. Then follows a slow decrease from around $7 R_{\odot}$ to $5 R_{\odot}$ for the Alfv\'en radius and from $3 R_{\odot}$ to $2 R_{\odot}$ for $r_{\mathrm{ss,opt}}$. This variation is in qualitative agreement with the variation suggested in \citet{Lee2011} and \citet{Arden2014}. 

It is interesting to see that the sonic point location is more steady over the cycle, located around $5.5 R_{\odot}$ (see Table \ref{table1}). Its location is mostly controlled by the thermodynamics we impose for our simulation and which remain constant in our entire sample. The sonic point is always located between the optimal source surface and the Alfv\'en point. The interaction between the magnetic field and the solar wind flow around the sonic point are crucial to understand to properties of the solar wind at 1 A.U \citep[see e.g.,][]{KoppHolzer1976,WangSheeley1991,Velli2010}, and we must investigate what is occurring in the higher corona, \textit{i.e.} beyond the optimal source surface. 

\section{Magnetic field organization in the higher corona}
\label{sec:pfssmhd}

\begin{figure*}
\center
\includegraphics[width=7in]{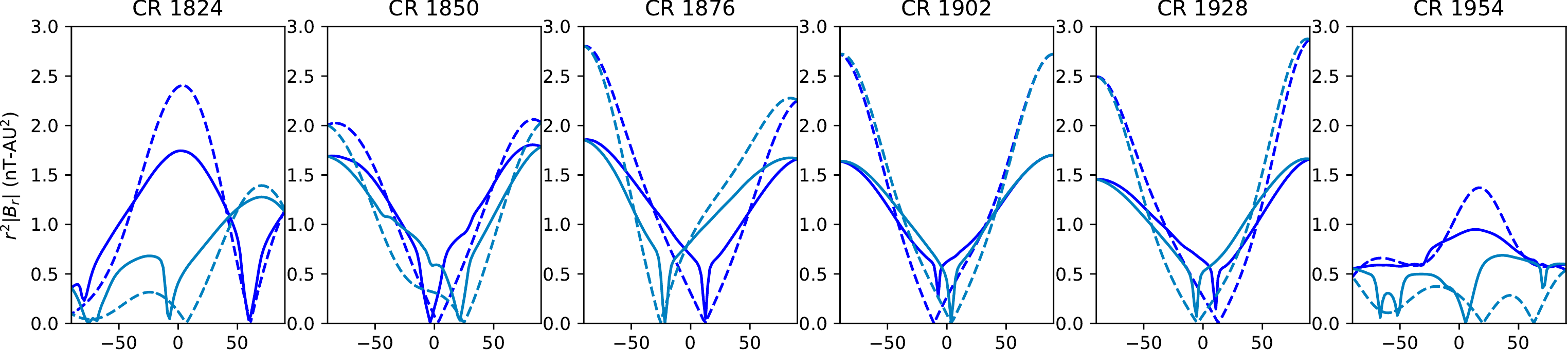}
\includegraphics[width=7in]{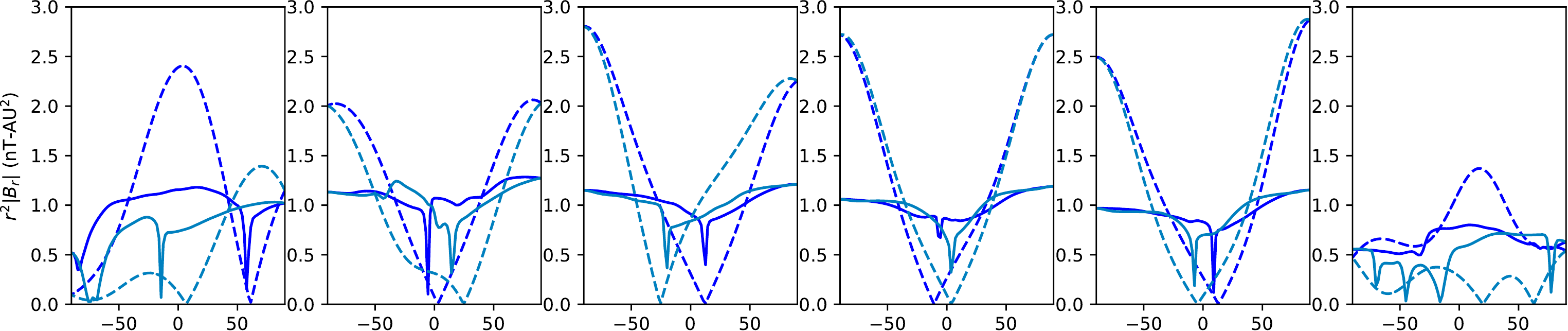}
\caption{Structure of the radial magnetic field on spheres of radii $3 R_{\odot}$ (upper panel) and $10 R_{\odot}$ (lower panel) at different epochs of the cycle as a function of latitude. Two longitudes ($\varphi=0,\pi$) are represented with the color gradation. In the upper panel, the structure of $B_r$ is relatively close to the a PFSS solution (computed at $r_{\mathrm{ss,opt}}$) shown in dashed lines. Further away, in the lower panel, $B_r$ is constant with latitude -which is in better agreement with Ulysses observations- and deviates significantly from the PFSS model.}
\label{Bflux}
\end{figure*}

In Figure \ref{Bflux} we show the radial magnetic field as a function of the latitude in the MHD model and the PFSS model for six Carrington rotations (same as in Figure \ref{AlfSurf}) at different heights: $3R_{\odot}$ and $10R_{\odot}$. Colors show two different longitudes ($\varphi=0,\pi$). The profiles are similar for epochs far from the maximum of activity when the magnetic field is globally axisymmetric, while they vary significantly near maximum. In the MHD model, at a height of $3 R_{\odot}$ -just above the optimal source surface-, the radial magnetic field varies strongly with latitude. Changes of polarity can be identified when $B_r  \rightarrow 0$, and we see $B_r$ decaying slowly on both sides of the current sheet, especially near minimum of activity. Higher in the corona, at $10 R_{\odot}$, the radial field forms a plateau at all latitudes but near current sheets where it decays extremely sharply. The standard deviation of $B_r(\theta)$ is about $10\%$ at $10 R_{\odot}$, against $33 \%$ at $3 R_{\odot}$ at activity minimum. This latitude independent radial magnetic field is in agreement  \textit{Ulysses} observations far away from the Sun and should be recovered in every MHD solar wind model.

This feature is not in general obtained with the PFSS model. In Figure \ref{Bflux}, radial magnetic field profiles of potential extrapolations made with the same surface magnetic field than the corresponding MHD models, and with the optimal source surface (see Table \ref{table1}) are shown with dashed lines. The PFSS model has, by definition, a fixed structure in space beyond the source surface. In both panel, the structure of the radial magnetic field is consequently the same: strongly modulated over latitude and maximal in the core of coronal holes. The standard deviation of $B_r(\theta)$ is in this model and for activity minimum around $66\%$.

\begin{figure}
\includegraphics[width=3in]{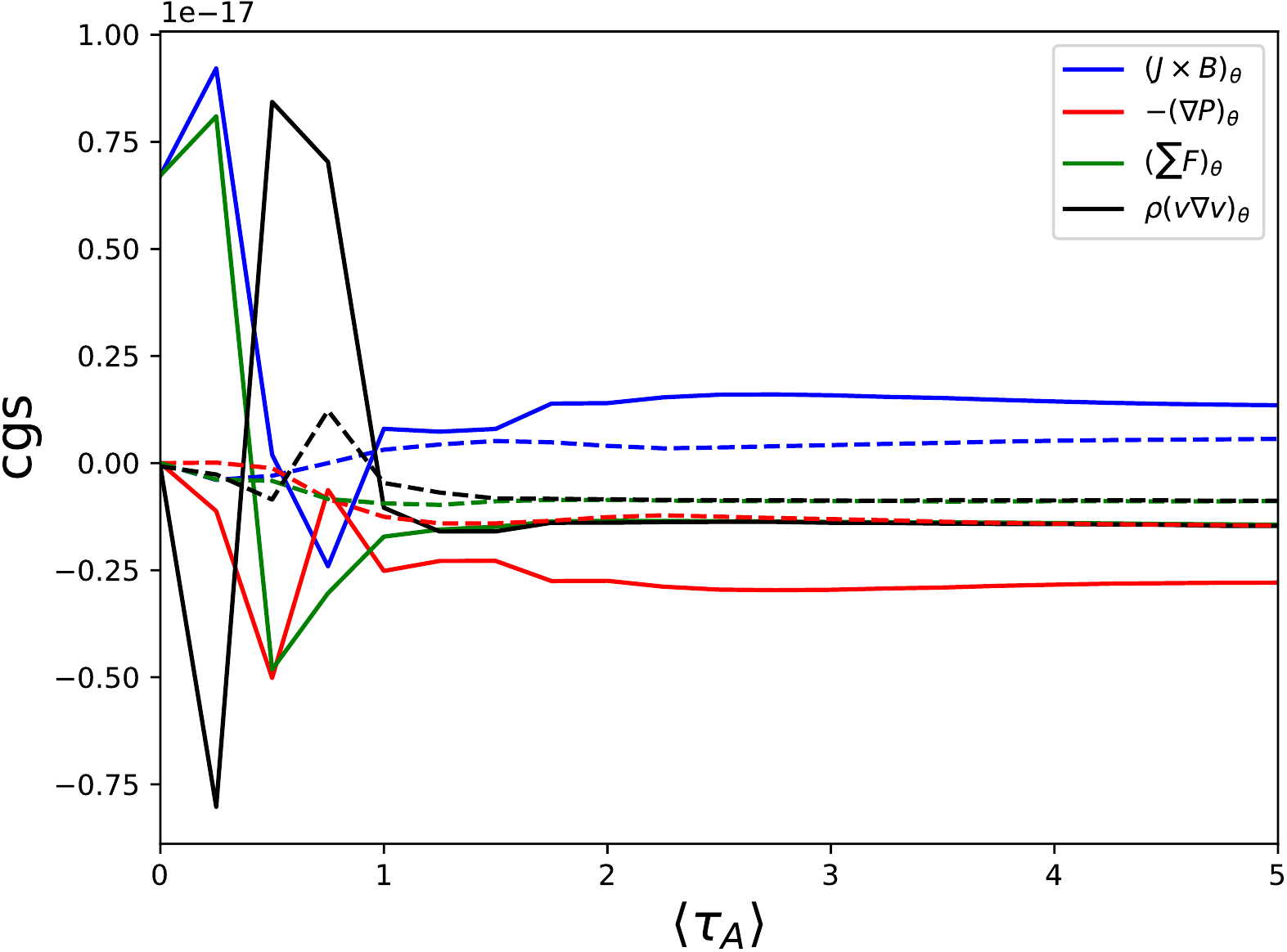}
\caption{Time evolution of the latitudinal terms in the momentum equation for a representative point of the domain, located in a coronal hole for CR 1902 at $r=10 R_{\odot}$, $\theta = \pi/4$, $\phi=0$. Time is expressed in average Alfv\'en crossing time in the domain $\langle \tau_A \rangle$. The plain lines represent the simulation initialized with $r_{\mathrm{ss,opt}}=2.3 R_{\odot}$. We see the initial latitudinal Lorentz force diminishing to reach equilibrium with the pressure gradient and the advection term. Dashed lines illustrate the simulation initialized with $r_{\mathrm{ss}}=15R_{\odot}$, which starts with no currents and thus no Lorentz force.}
\label{ForceEvol}
\end{figure}

\citet{SmithBalogh1995} discussed this feature and noted that any latitudinal gradient of $B_r$ existing in the corona would tend to vanish through the action of the latitudinal term of the Lorentz force (created by the azimuthal current). It is worth noting that, in general, the PFSS model does not yield a force free or even a current free magnetic field beyond the source surface\footnote{By definition, the field is current free below the source surface.}. Given the case of a pure axisymmetric dipole, the latitudinal term in the Lorentz force beyond the source surface can be written:

\begin{eqnarray}
\left (\mathbf{J} \times \mathbf{B} \right)_{\theta} &= & J_{\varphi} B_r - J_r B_{\varphi},\\ 
&=& - \frac{1}{r} \frac{\partial B_r}{\partial \theta} B_r, \\
&=&\frac{1}{r^5} f(r_{\mathrm{ss}})^2 \cos(\theta) \sin(\theta),
\end{eqnarray}
where $f(r_{\mathrm{ss}})= 3\mu /(2 r_{\mathrm{ss}} + 1/r_{\mathrm{ss}}^2)$, and $\mu$ is the dipole moment. The $\theta$ component of the Lorentz force is thus always equatorward. Put in other words, the gradient of $B_r(\theta)$ shown in Figure \ref{Bflux} for the PFSS model would create a latitudinal magnetic pressure and disappear if left to evolve. To illustrate this process, we initialize a simulation of CR 1902 at the optimal source surface, and we follow the evolution of the latitudinal term of the momentum equation for point located at a height of $10 R_{\odot}$, until steady state is reached. Figure \ref{ForceEvol} shows the time evolution of the different terms.

When the simulation is initialized with $r_{\mathrm{ss,opt}}$, the initial latitudinal Lorentz term is the dominant force and decreases to reach equilibrium with the pressure gradient and advection terms. When the simulation is initialized with a large $r_{\mathrm{ss}}$, the Lorentz force increases slightly but reaches a lower steady-state value than in the latter case. Note that the two steady-state are not exactly equivalent in Figure \ref{ForceEvol}. The latitudinal profile of $B_r$ is less flat, especially near the poles, when the simulation is initialized with a small $r_{\mathrm{ss}}$. Also, the surface (boundary) $B_{\theta}$ field is a function of the source surface radius. The dependent part, however, quickly tends towards zero when $r_{\mathrm{ss}}$ tends towards infinity. We would thus advise to start such kind of MHD simulations with a large initial value of the source surface radius, at least larger than the zone of interest. 

\begin{figure}
\includegraphics[width=3.3in]{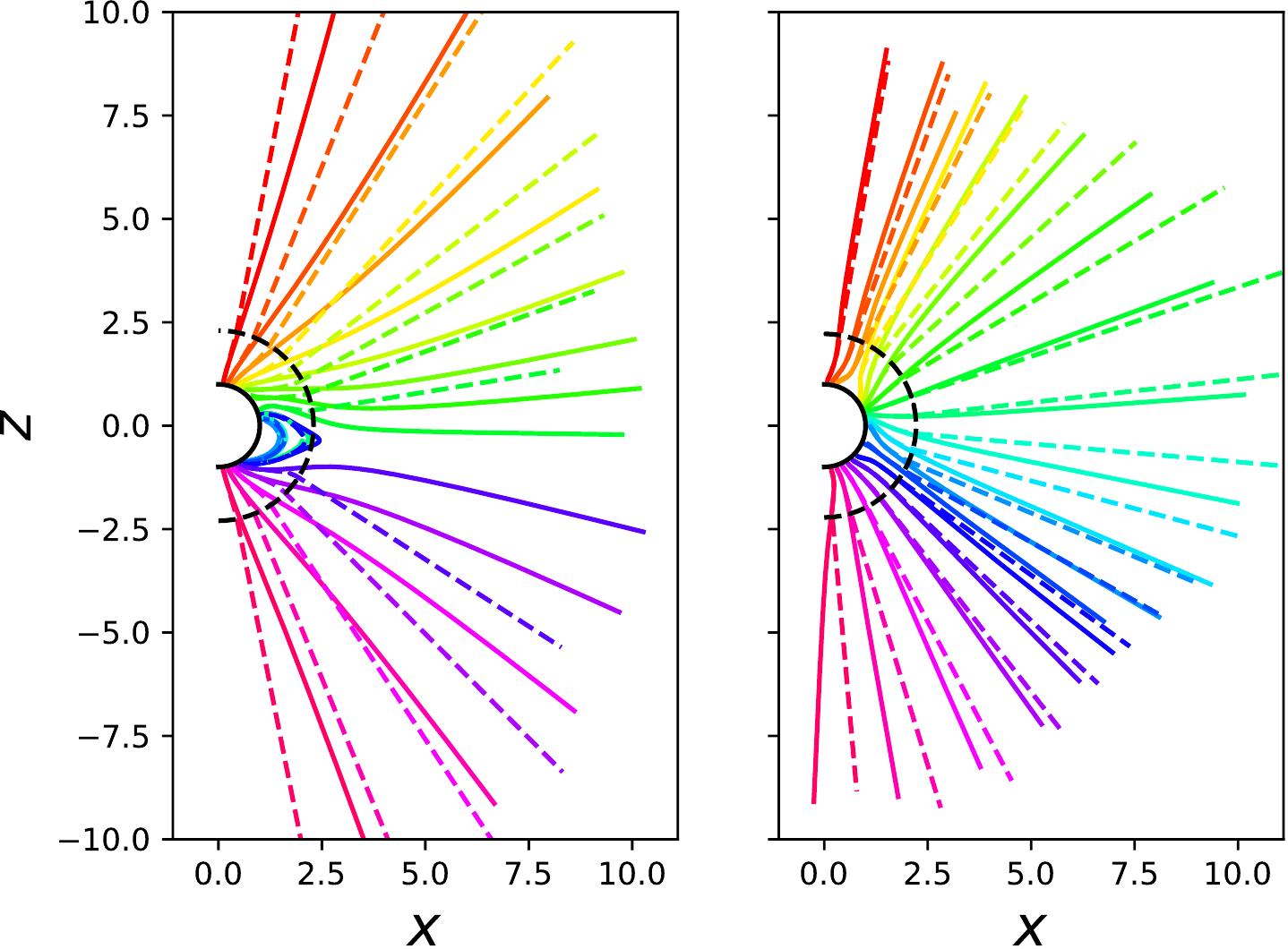}
\caption{Difference between the magnetic field lines of a MHD simulation and a PFSS model at the optimal source surface for minimum of activity (left panel, CR 1902) and maximum of activity (right panel, CR 1954). For both models, field lines departing from the same footpoint at the base of the corona are identified by a single color in the two models. We see that field lines diverge approximately at the optimal source surface (shown in dashed lines) at both epochs. Moreover the difference is more important during minimum, because of the more significant reorganization of the field lines that occurs beyond $r_{\mathrm{ss,opt}}$.}
\label{MHD_PFSS}
\end{figure}

In Figure \ref{MHD_PFSS} we show the difference between the geometry of a given flux tube integrated from the surface in a PFSS solution and in the MHD simulations for two epochs (minimum and maximum of activity). As in Figure \ref{Bflux}, the PFSS is computed with the optimal source surface (see Table \ref{table1}) and thus both models have exactly the same open magnetic flux. Up to the optimal source surface radius, the field line structure is alike in both models \citep[see][]{Reville2015b}, confirming that the PFSS is a good proxy for the structure of closed loops, and the overall connectivity of the heliospheric magnetic field \citep{Riley2006}. The difference between the two models becomes significant beyond the source surface. Near solar minimum (left panel), the field lines of the MHD model are clearly closer to the current sheet (or to the equator), which one could cautiously call a "equatorward migration" in the MHD model, consistent with the analysis above.

Differences are also visible during maximum of activity, and the same phenomenon occurs in each and every coronal hole. However the deviation from PFSS seems less significant at maximum of activity. Then, higher order multipoles -notably a strong quadrupolar component- make the magnetic pressure decaying more rapidly in the solar atmosphere. The outflow is thus able to make the necessary adjustments lower in the corona. The radial field is homogeneous closer near maximum, which can be observed in Figure \ref{Bflux}, particularly for CR 1954.

\section{The wind speed / expansion correlation over the cycle}
\label{sec:wsc}

\begin{figure}
\center
\includegraphics[width=3.5in]{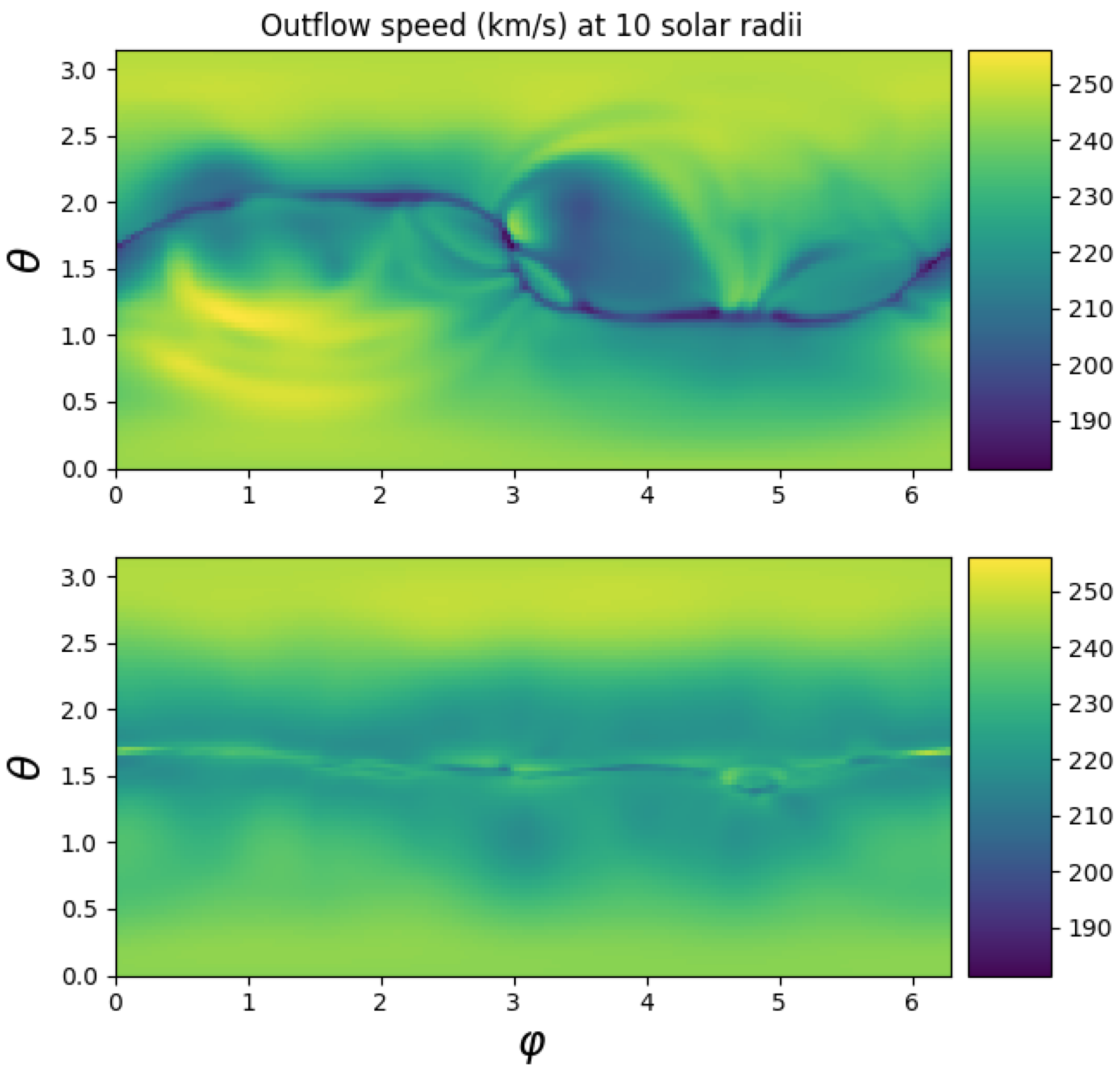}
\caption{$(\theta,\phi)$ maps of the radial velocity at $10 R_{\odot}$ for CR 1837 (activity maximum, upper panel) and CR 1902 (activity minimum, lower panel). The complex structure of the magnetic field shapes the speed distribution at maximum of activity.  At minimum of activity, the velocity structure is globally axisymmetric with slower winds at the equator and faster winds within polar coronal holes.}
\label{vrmap}
\end{figure}

The continuous expansion that occurs at various heights in the corona must have an influence on the wind speed. Figure \ref{vrmap} shows the spatial distribution of the radial wind speed at height of $10 R_{\odot}$ for CR 1837 (maximum of activity) and CR 1902 (minimum of activity). Due to the absence of Alfv\'en wave induced mechanisms, the maximum wind speed at $10 R_{\odot}$ is around $250$ km/s while slowest winds reach two thirds of this value. At the edge of the simulation domain, the fast wind goes beyond $400$ km/s and is close to the terminal speed given by a 1D polytropic model. Near activity maximum, slow and fast wind streams are located at any latitude, and we can identify the so-called ballerina skirt that forms an S shape along all longitudes. For activity minimum, the global axisymmetry of the magnetic field structure is found in the speed distribution, with faster streams at high latitude, and slower streams near the equator.

Following the observations of \citet{WangSheeley1990}, the terminal wind speed of the solar wind is generally thought to be anti-correlated to the expansion factor, which computes the deviation from a purely radial magnetic field and which can be defined as:
\begin{equation}
f_{\mathrm{exp}} (s)= \frac{B_0}{B(s)} \left(\frac{r_0}{s} \right)^2,
\end{equation}
where $r_0$ is the radius where the field lines integration starts (here just above the surface), and $s$ is the curvilinear abscissa along the field line. More specifically, this expression corresponds to the total expansion factor, and will measure the accumulated expansion along the field line. As soon as the field becomes purely radial, the total expansion factor becomes constant.

\begin{figure*}
\begin{tabular}{cc}
\includegraphics[width=3.3in]{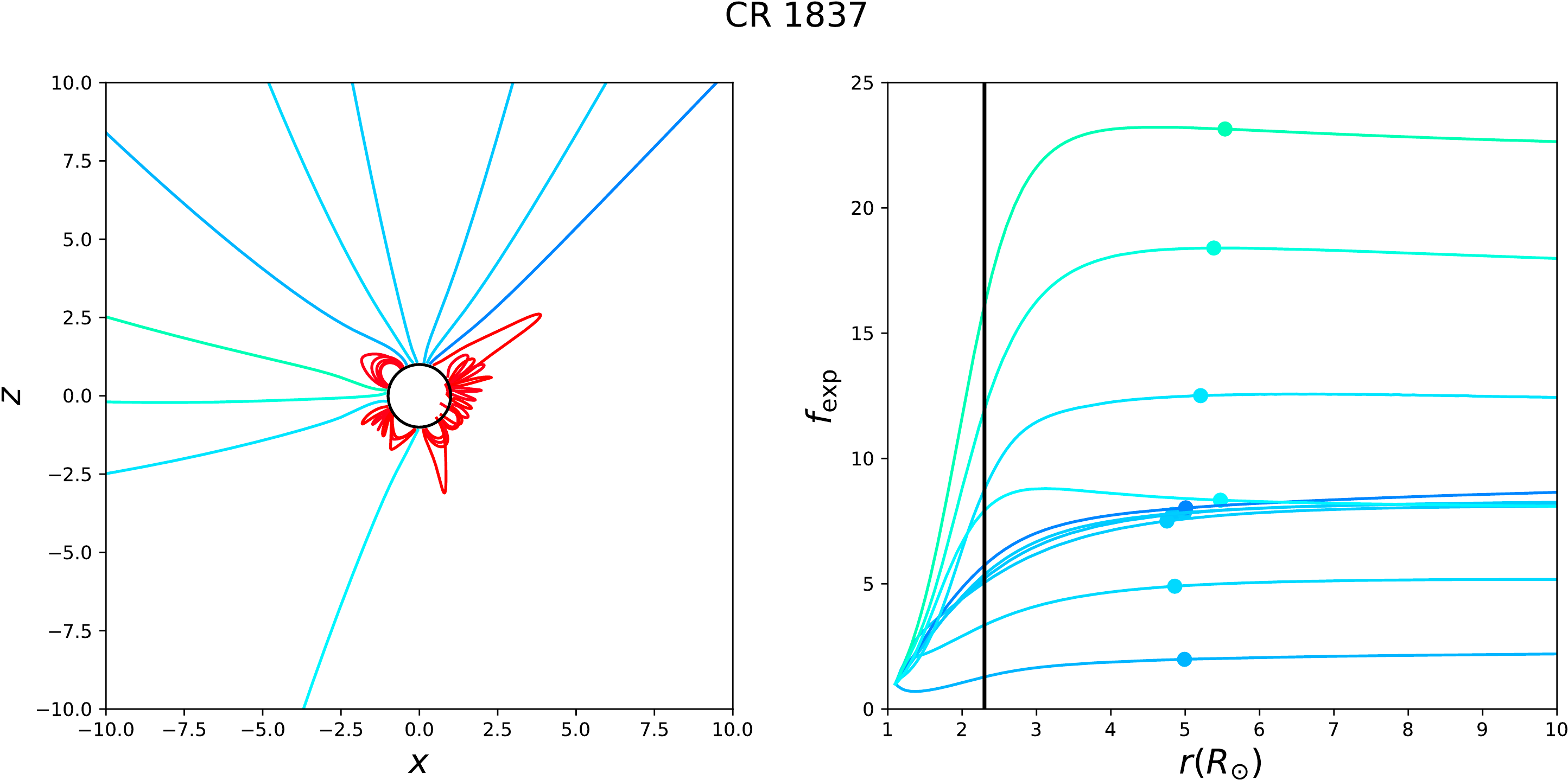} & \includegraphics[width=3.3in]{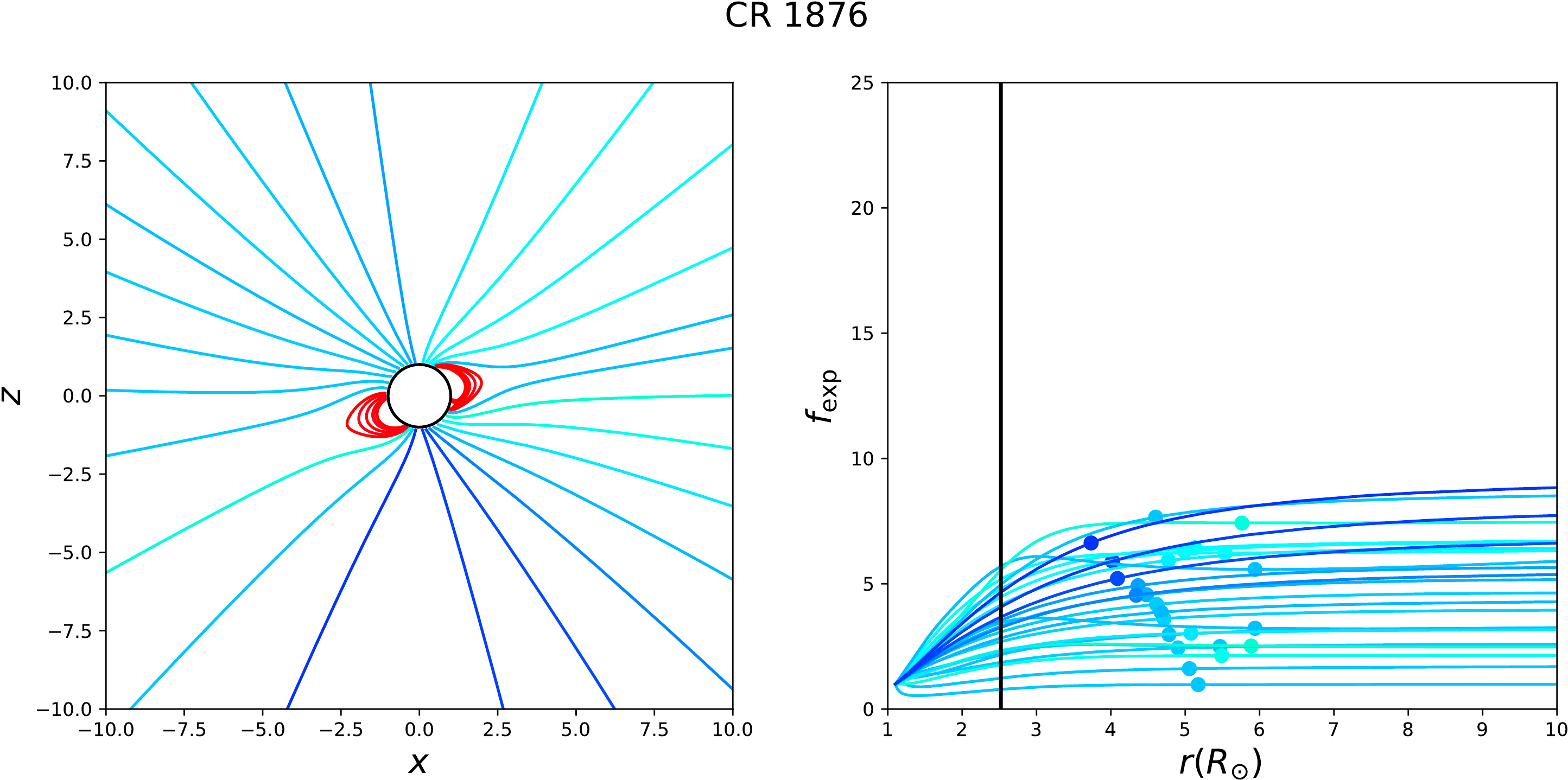} \\
\includegraphics[width=3.3in]{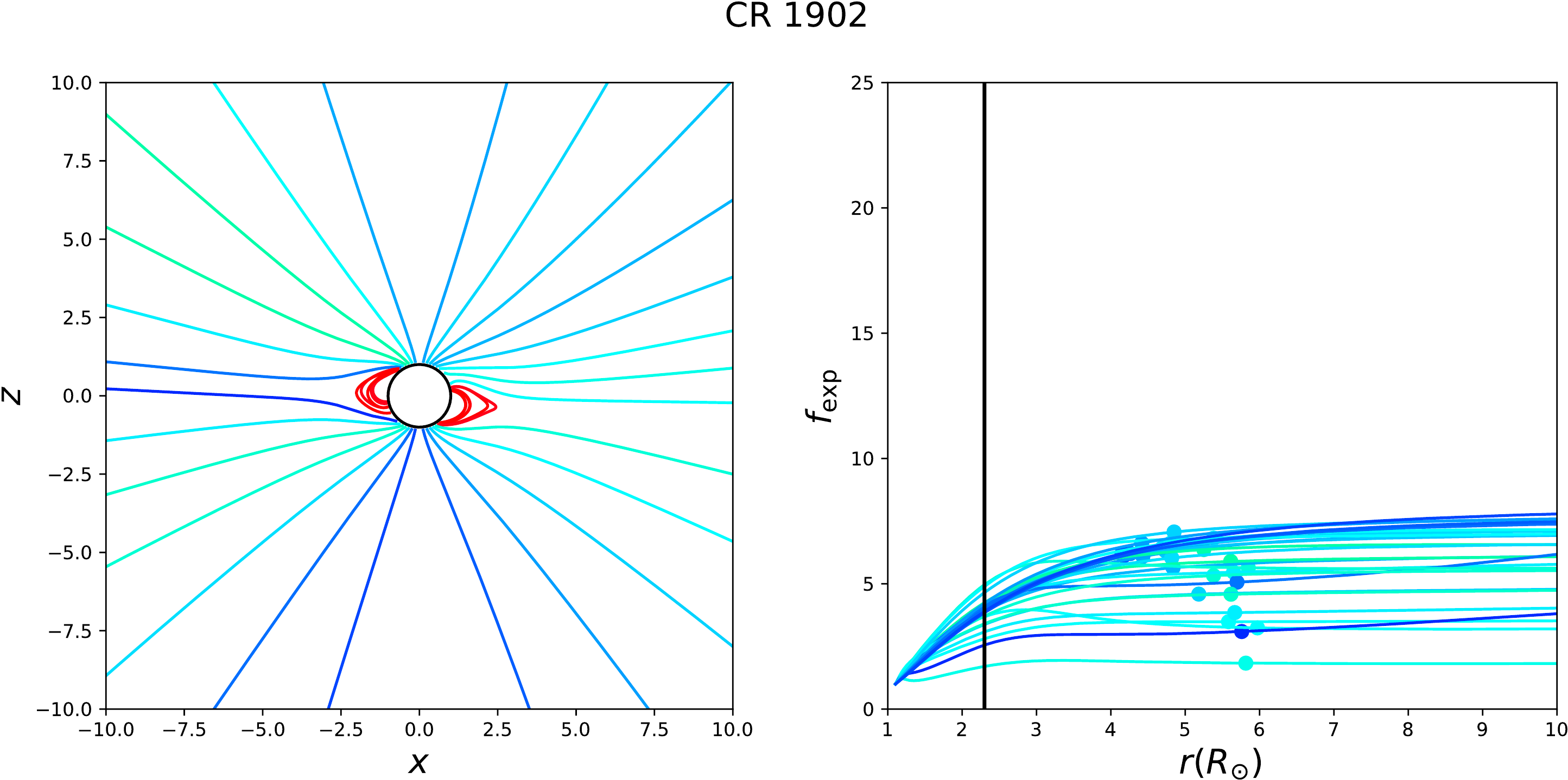} & \includegraphics[width=3.3in]{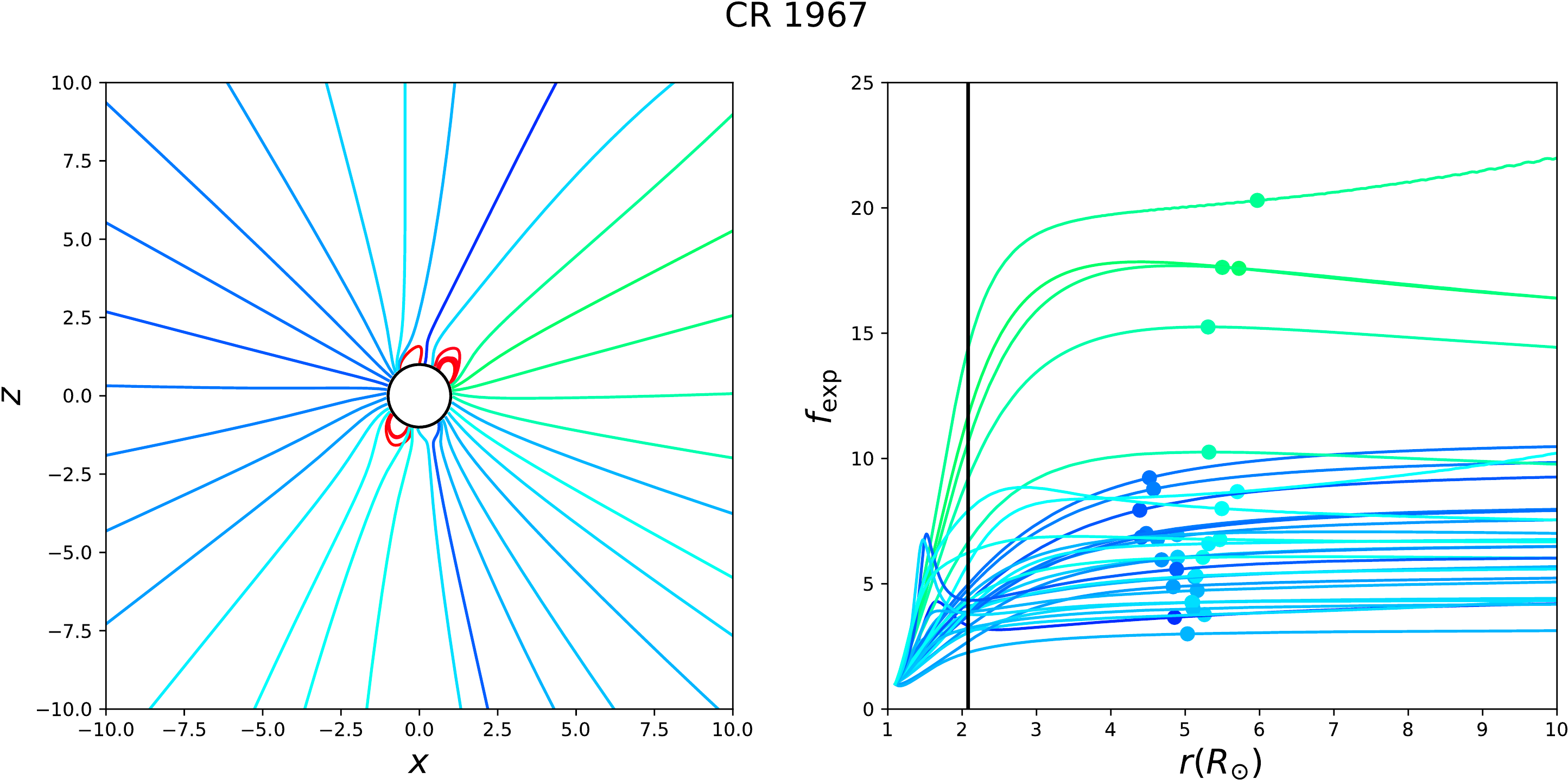} 
\end{tabular}
\caption{Field lines projections on the $(x,z)$ plan (left panel) and profile of the of the expansion factor as a function of $r$ (right panel) for Carrington rotations 1837, 1902, 1876, 1967. Colors are associated with the velocity of the outflow along the field lines at $10 R_{\odot}$. Red field lines correspond to coronal loops and never reach this distance from the star. Greener field lines show slower solar wind, and bluer field lines represent faster winds. Slowest winds are usually flowing along coronal loops. The total expansion reach higher levels near maximum (CR 1837, 1967), and in all cases goes on beyond $r_{\mathrm{ss,opt}}$  (black vertical line) and the sonic point (shown as a same color point on the expansion factor profiles).}
\label{fexpCycle}
\end{figure*}

In Figure \ref{fexpCycle}, we show the structure of 20 field lines for four Carrington rotations, and the total expansion factor for open field lines among them. In all the cases shown in Figure \ref{fexpCycle} the expansion profile is still evolving at 10 solar radii, \textit{i.e.} well beyond $r_{\mathrm{ss,opt}}$, which is shown with the black vertical line and defined non-locally thanks to a procedure based on the open flux of the simulation (see section \ref{sec:num}). As shown in Table \ref{table1}, the expansion is also acting beyond the sonic point (defined for each field lines and superimposed with the same color code on the expansion profile in Figure \ref{fexpCycle}) and the Alfv\'en point. 

The colors in Figure \ref{fexpCycle} are associated with the wind speed at $10 R_{\odot}$, and goes from light green (slow wind) to to dark blue (fast wind). As expected, slower winds are generally flowing along coronal loops (whose field lines are colored red), and are thus located near the dipolar equator at activity minimum (CR 1876 and 1902), and at many different latitudes during maximum (the projection in the 2D plane may over simplify near maximum configurations where small coronal loops exist at all latitudes). The total expansion factor reaches higher values near maximum (around $25$) than near minimum (where it stays below $10$). Highest total expansion factors are generally associated with a sharp increase of $f_{\mathrm{exp}}$ below the optimal source surface and relatively constant profiles beyond $4 R_{\odot}$. This matches the classical description of the interaction between the expansion and the flow. When located below the sonic point, super radial expansion (= increasing $f_{\mathrm{exp}}$), increases the mass flux and decelerates the wind \citep[see][]{WangSheeley1991,Velli2010}, as less energy is available -per particle- to be converted in bulk kinetic energy when the wind reaches its terminal speed. 

\begin{figure*}
\center
\includegraphics[width=7in]{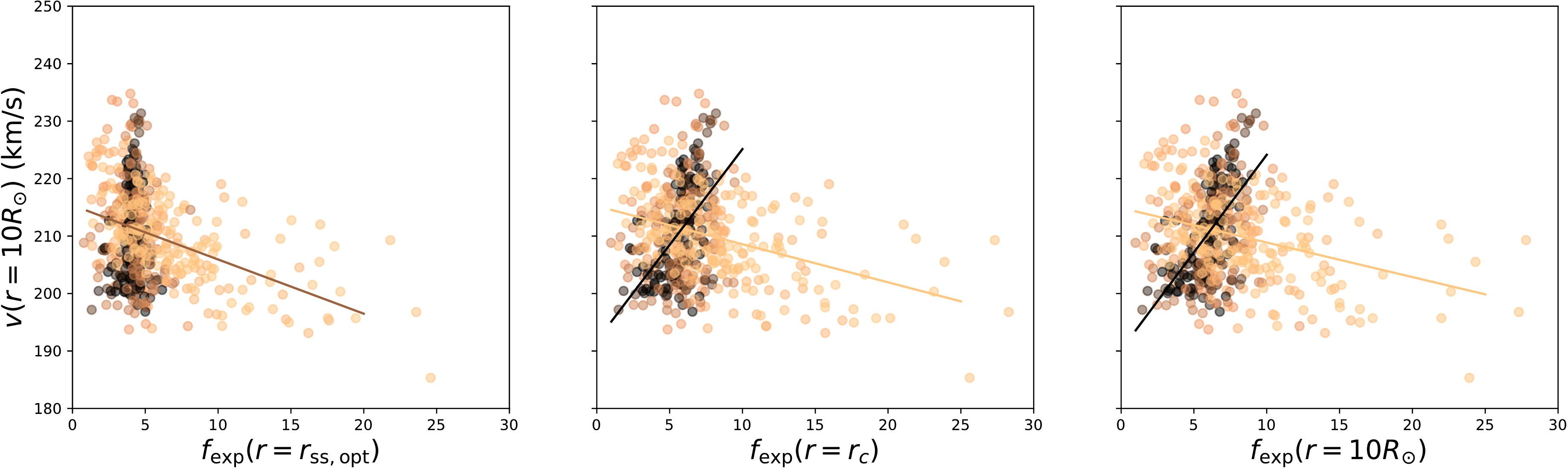}
\includegraphics[width=4in]{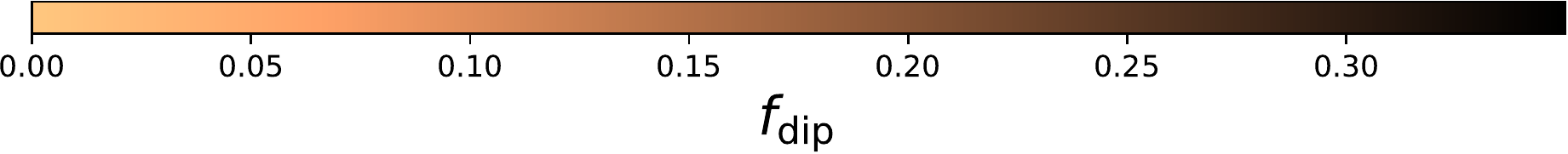}
\caption{Correlation between the wind speed and the expansion factor at three different heights in the corona: the optimal source surface radius (on the left panel), at the sonic point (middle panel) and at $10 R_{\odot}$ (where the wind speed is computed for all panels). The colors indicate the parameter $f_{\mathrm{dip}}$, so that orange colors indicate a very weak dipolar component (near maximum) and black colors indicate a strong dipolar component (near minimum). We see that the correlation changes with the cycle phase. The positive correlation between $f_{\mathrm{exp}}$ and the wind speed is the strongest at large heights, due to expansion beyond the sonic point.}
\label{VvExp}
\end{figure*}

However, this anti-correlation is not observed near minimum of activity in the higher corona. CR 1876 and 1902 in Figure \ref{fexpCycle} seem to show the opposite trend where blueish expansion profiles cross greener ones with increasing distance. This feature becomes clear above the sonic point, where the continuous super radial expansion does accelerate the solar wind \citep[because the wind is supersonic and no more able to pump material from the surface, see, e.g.][]{Reville2016csss}. As shown in section \ref{sec:pfssmhd}, within large coronal holes, the magnetic field is significantly expanding beyond the surface in order to homogenize its radial component. The acceleration beyond the sonic point seems to be the dominant process that determines the relation between the total expansion factor and the wind speed at $10 R_{\odot}$. 

The same process occurs in all coronal holes, even when the field geometry is more complex, near maximum of activity. However, in this configuration, the expansion around small closed loops is much more important, making the expansion factor at the source surface actually the most relevant parameter to determine the wind speed. Figure \ref{VvExp} deepens this analysis. We show the wind speed at $r=10 R_{\odot}$ as a function of the total expansion factor computed for the 40 open field lines (among which are the one of Figure \ref{fexpCycle}) at three different heights: the optimal source surface, the sonic point, and at $10$ solar radii. The scattered points are colored according to the parameter $f_{\mathrm{dip}}$, which is a good proxy for the cycle phase (see Figure \ref{scmap}). Considering all points as a whole, we observe a global anti-correlation between the expansion factor and the wind speed, which is in agreement with observations \citep[see][]{WangSheeley1990,Wang2016}. At the source surface, the global anti-correlation is shown with a brownish line, and no correlation can be extracted for the point near minimum alone.

Nevertheless, higher in the corona, we can obtain two different regressions, for simulations whose $f_{\mathrm{dip}}$ parameter is over or below $50\%$ of the maximum value ($f_{\mathrm{dip,max}}=0.35$) for the period considered. For near minimum cases, we get a positive correlation (shown by the black line) between the wind speed and the expansion factor starting at the sonic point. This correlation gets better as we go up in height at $10 R_{\odot}$. Hence, for simple, near dipolar, configuration of the solar magnetic field, where most of the open field lines come from the large polar coronal holes, the expansion around and beyond the sonic point operates a significant reorganization. Of course, beyond the source surface, the total area is conserved and both super and under radial expansion must compensate. As such, super (under) radial expansion increases (decreases) the wind speed, and the correlation between the terminal speed and the total expansion factor is reversed \citep[see also][]{Pinto2017}. The additional acceleration provided only by the super radial expansion is maximum on field lines that cross local maxima of $B_r(\theta,r_{\mathrm{ss,opt}})$. This process must influence the acceleration of the fast solar wind, which is thought to occur at large distances from the Sun, where Alfv\'en waves dissipate. The correlation (shown by the orange line) stays globally negative for points near maximum of activity everywhere in the corona.

\begin{figure}
\center
\includegraphics[width=3.2in]{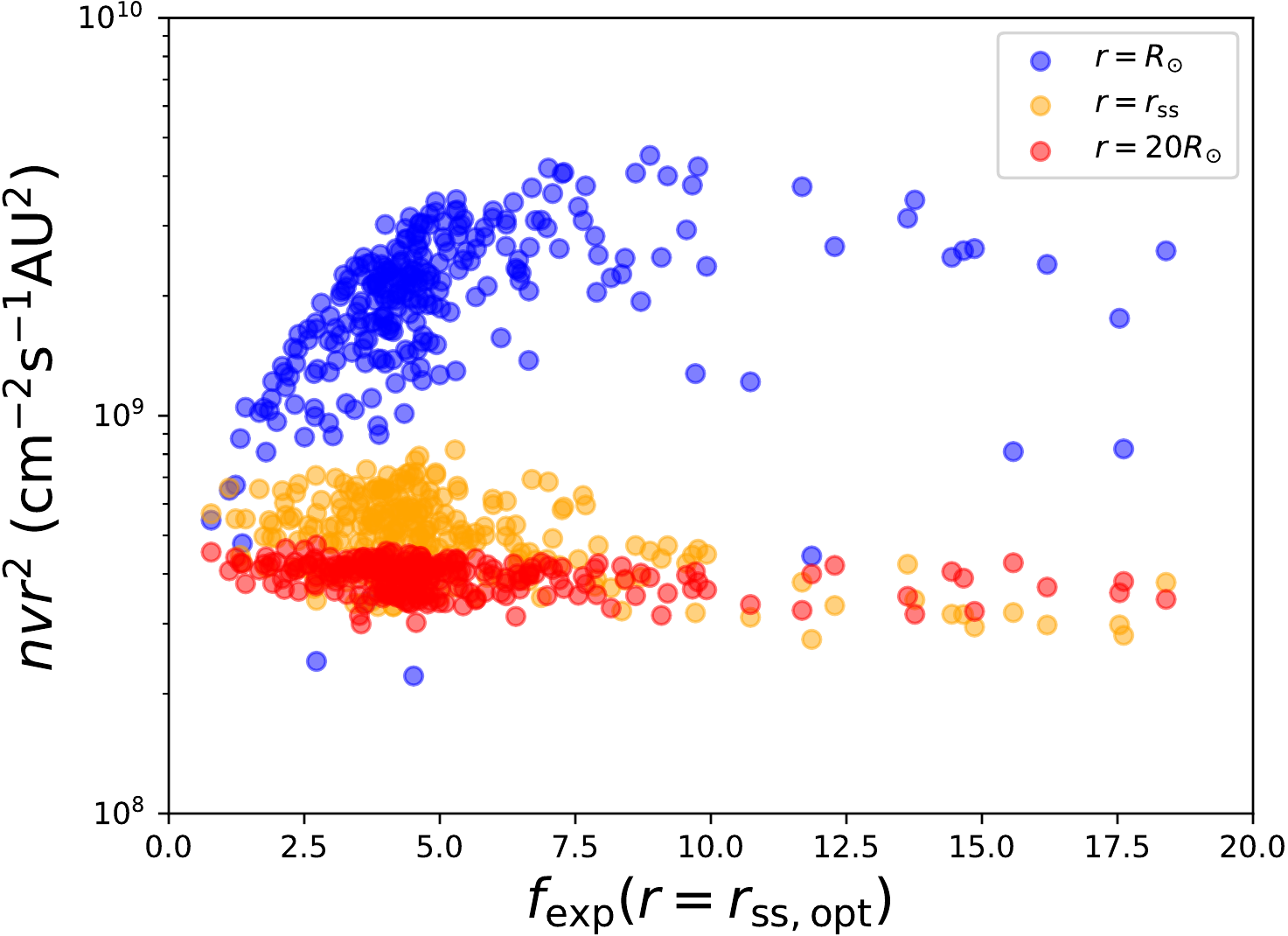}
\caption{Mass flux (or proton flux, here $n=\rho/m_p$) represented at three different heights as a function of the expansion factor in the low corona ($r=r_{\mathrm{ss,opt}}$). Blue dots represent the mass flux at the solar surface, orange dots the mass flux at the source surface and red dots the mass flux at a height of $20 R_{\odot}$. We see that $n v r^2 (R_{\odot})$ and $f_{\mathrm{ss}}$ are positively correlated. The mass flux tends toward a mostly constant profile higher in the corona, with help of the super-radial expansion beyond $r_{\mathrm{ss}}$.}
\label{rhovexp}
\end{figure}

In Figure \ref{rhovexp}, we show the mass flux as a function of the expansion factor at the source surface $f_{\mathrm{ss}}$) at different heights in the corona. As shown with the blue dots, a large expansion below the source surface (associated with slow streams) increases the mass flux at the coronal base, through the velocity that is left free in the boundary. The mass flux at the surface spans over an order of magnitude, and the mass loading diminishes the energy per particle and the terminal speed. This explains the good correlation between the wind speed and the expansion factor at the source surface. However, near the source surface, the mass flux/expansion correlation is reversed with faster wind (low $f_{\mathrm{ss}}$) having a larger mass flux. The standard deviation from the mean is $22$\% at $r=r_{\mathrm{ss}}$. The latter correlation is not observed in the solar wind where the mass flux is almost constant \citep[although slightly larger in the slow solar wind, see][]{Holzer2005}. Expansion above the source surface acts to correct this relationship decreasing the mass flux in faster winds components (low $f_{\mathrm{ss}}$). This is what we observe going up in height at $20 R_{\odot}$ (red points), where we obtain a rather constant profile of the proton flux around $4 \times 10^8$ cm$^{-2}$s$^{-1}$ with a standard deviation of $8$\%, more in agreement with observations.

In the work of \citet{Wang2010}, the relative constancy of the mass flux is explained through a linear relation between the mass flux at the surface and magnetic field strength at the coronal base as well as the energy input from the chromosphere. Indeed, rapid expansion below $2.5 R_{\odot}$ can lead to an enhanced local deposition of heat in the low corona, which helps lifting material from the chromosphere as shown by \citet{Wang2009}. Although the heating mechanisms used in this paper are not related to the magnetic field at the coronal base, we also get higher mass flux for higher surface field strength. This can be understood by the correlation between high expansion in the low corona and high magnetic flux concentration at the surface associated with small scale structures. Our model thus agree with more elaborated thermodynamical model of the corona. Also, the fact that we keep the coronal temperature fixed at the base of the corona help the simulations comply with this observation \citep[because the mass flux depends exponentially on the coronal temperature, see e.g.][]{HansteenVelli2012}. Reaching a quantitative agreement with solar wind observables needs both an accurate description of the heating in the low corona and of the continuous expansion up to tens of solar radii \citep[see e.g.][where the expansion in higher corona has been added to a multi flux tube model based on a PFSS]{Pinto2017}.

\section{Discussions}
\label{sec:disc}

In this study, we have shown that the varying structure of the surface solar magnetic field over the 11-yr cycle has a strong influence on the higher corona, and the organization of the interplanetary magnetic field. Using fixed thermodynamics, we showed that the characteristic surfaces of the solar corona varies over the cycle by $20-30\%$ at most, which is less than what others works have suggested \citep[see][]{PintoBrun2011}. The optimal source surface radii, that contains all coronal loops, are as expected between $2$ and $3$ solar radii. The sonic point is located between $5 R_{\odot}$ and $6 R_{\odot}$, while the Alfv\'en point is located slightly further on. Taking into account both the organization of the field (the topology) and its energy (or surface flux/strength) is crucial to obtain these results. Both seem to balance over the cycle to limit variations of the Alfv\'en surface. Although more realistic heating prescriptions are likely to have a strong influence on the location of the critical surfaces, and their variation over the cycle, a similar result has been obtained with 3D MHD simulations for two epochs (minimum and maximum) with an Alfv\'en wave turbulence heating in the work of \citet{Garraffo2016}. This balance may be a reliable output of the dynamo mechanisms at the source of the solar magnetic field.

The constant radial field in latitude observed by \textit{Ulysses}, is the result of an equilibrium between magnetic forces and flow dynamics, which is occurring at different heights over the cycle. The PFSS model, which imposes an oversimplified structure of the corona beyond $2.5 R_{\odot}$, cannot capture this process. Moreover, because the magnetic pressure decays strongly close to the star near maximum of activity, this equilibrium is met closer than during minimum of activity. In terms of field expansion, this translates to an extended expansion in the higher corona, more significant near minimum.  As a consequence, depending on the heights considered for the total expansion factor, the net effect on the wind velocity can be reversed between the low corona (at the source surface) and the higher corona (here around $10 R_{\odot}$) for certain period of the cycle, characterized by a strong dipolar component. 

This reorganization influences the solar wind properties by increasing the speed and decreasing the mass flux in the core of coronal holes, where $B_r(\theta)$ is maximum at the source surface. It is also necessary to get a mass flux relatively independent of the wind speed at large distances. In the theoretical model of \citet{WangSheeley1991}, used to explain the observations of \citet{WangSheeley1990}, a non-linear relationship between the expansion factor at the source surface $f_s$, or the sonic point $f_c$, and the total expansion at Earth orbit $f_E$ is introduced. The relationship reads $f_s \approx f_c = 3 f_E^{1/2}$, which means that whenever $f_c$ is less than $9$, the total expansion will increase between the sonic point and the observer, thus accelerating the wind and reducing the mass flux. For $f_s > 9$, the opposite will occur, yielding a rather constant mass flux for all wind speed. The transition value used in the relation of \citet{WangSheeley1991} matches the expansion limit we observe at minimum of activity, and confirms that the expansion parametrized in \citet{WangSheeley1991} is due to the pressure equilibrium met in the higher corona at solar minimum.

Hence, despite simple driving mechanisms, our model renders a general process that should be taken into account in every more sophisticated solar wind model. Moreover, our study might have consequences on the interpretation of the WSA semi-empirical relationship \citep[see][]{ArgePizzo2000,Arge2003}. In this formulation, another parameter $\theta_b$ is used to control the terminal wind speed. It represents the minimum angular distance to a coronal hole boundary and is thought to be strongly correlated to $f_s$ \citep{McGregor2011}, so that fast wind streams emerge from deep within coronal holes while slow wind originates near the streamers. We have shown that within large coronal holes (large $\theta_b$), the expansion beyond the source surface has a strong influence on the wind speed. It is likely that part of the efficiency of this parameter is due to what happens in the upper corona rather than near photosphere processes. A proper description of the expansion below and beyond the sonic point over an extended and realistic profile of a solar wind flux tube will certainly provide more accurate and physically motivated predictive laws of the wind terminal speed. 

As demonstrated by many works \citep[see e.g.][]{Leer1982, Withbroe1988}, an extended energy deposition is needed to reproduce the fast solar wind. A more detailed thermodynamics is key to achieve a fully consistent picture, and recent work have introduced advanced Alfv\'en wave related pressure and heating in 3D MHD simulations \citep{Lionello2009,Sokolov2013,vanDerHolst2014}. However, we believe that additional heating and momentum deposition will not qualitatively change the results we get in this paper. As stated earlier, the model of \citet{WangSheeley1991} -which includes these processes- needs the high corona reorganization described in this work. More realistic models will likely get the sonic point closer to the Sun, near the optimal source surface radius in open regions (fast solar wind). Because the PFSS model is able to relatively precisely recover the structure of the corona up to $2.5R_{\odot}$, the observed homogenization of the radial magnetic field must occur beyond. Thus, beyond the sonic point where Alfv\'en waves are thought to be dissipated, the influence of expansion on the solar wind speed may be even more important than presented here.

Our work could also contribute to a more accurate description of the expansion in local models that focus on a given flux tube. This models are now, the only models able to capture a fully self-consistent description of the propagation and the dissipation of Alfv\'en waves \citep{SuzukiInutsuka2006,Cranmer2007,MatsumotoSuzuki2012}. Yet they generally use fast and very localized expansion prescriptions of what happens below the source surface \citep[see the general formulation of][]{KoppHolzer1976}. Improving this kind of models and comparing them to observations will help better understand the acceleration profile of the solar wind in open regions. The upcoming spacecraft Parker Solar Probe \citep{FoxPSP2016} will continuously measure the evolution of the wind speed and the mass flux down to $9R_{\odot}$. Solar Orbiter \citep{Muller2013}, traveling out of the ecliptic, will be the successor of \textit{Ulysses} while approaching the Sun much closer. Both satellites will probe the nature and the location of the different mechanisms addressed in part in this paper, that are responsible for the solar wind observables at 1 AU.

\section{Acknowledgements}
The authors are grateful to A. Mignone and the PLUTO development team. We thank M. DeRosa for providing the processed observational data from the Wilcox magnetograms, R. Pinto, A. Rouillard, and M. Velli for valuable discussions and the anonymous referee for his/her comments that significantly enriched the discussion. We acknowledge funding by the ERC STARS2 207430, PNST and CNES via Solar Orbiter. High performance computations were performed on the machine Turing (IDRIS) within the GENCI 1623 program.


\begin{thebibliography}{59}
\providecommand\natexlab[1]{#1}
\providecommand\JournalTitle[1]{#1}

\bibitem[{{Altschuler} \& {Newkirk}(1969)}]{AltschulerNewkirk1969}
{Altschuler}, M.~D., \& {Newkirk}, G. 1969,
  \href{http://dx.doi.org/10.1007/BF00145734}{\JournalTitle{\solphys}, 9, 131}

\bibitem[{{Arden} {et~al.}(2014){Arden}, {Norton}, \& {Sun}}]{Arden2014}
{Arden}, W.~M., {Norton}, A.~A., \& {Sun}, X. 2014,
  \href{http://dx.doi.org/10.1002/2013JA019464}{\JournalTitle{Journal of
  Geophysical Research (Space Physics)}, 119, 1476}

\bibitem[{{Arge} {et~al.}(2003){Arge}, {Odstrcil}, {Pizzo}, \&
  {Mayer}}]{Arge2003}
{Arge}, C.~N., {Odstrcil}, D., {Pizzo}, V.~J., \& {Mayer}, L.~R. 2003,
  \href{http://dx.doi.org/10.1063/1.1618574}{in American Institute of Physics
  Conference Series, Vol. 679, Solar Wind Ten, ed. M.~{Velli}, R.~{Bruno},
  F.~{Malara}, \& B.~{Bucci}}, 190

\bibitem[{{Arge} \& {Pizzo}(2000)}]{ArgePizzo2000}
{Arge}, C.~N., \& {Pizzo}, V.~J. 2000,
  \href{http://dx.doi.org/10.1029/1999JA000262}{\JournalTitle{\jgr}, 105,
  10465}

\bibitem[{{Balsara} \& {Spicer}(1999)}]{BalsaraSpicer1999}
{Balsara}, D.~S., \& {Spicer}, D.~S. 1999,
  \href{http://dx.doi.org/10.1006/jcph.1998.6153}{\JournalTitle{Journal of
  Computational Physics}, 149, 270}

\bibitem[{{Cohen}(2015)}]{Cohen2015SoPh}
{Cohen}, O. 2015,
  \href{http://dx.doi.org/10.1007/s11207-015-0739-3}{\JournalTitle{\solphys}},
  \href{http://arxiv.org/abs/1507.00572}{{\sffamily arXiv:1507.00572
  [astro-ph.SR]}}

\bibitem[{{Cranmer} {et~al.}(2007){Cranmer}, {van Ballegooijen}, \&
  {Edgar}}]{Cranmer2007}
{Cranmer}, S.~R., {van Ballegooijen}, A.~A., \& {Edgar}, R.~J. 2007,
  \href{http://dx.doi.org/10.1086/518001}{\JournalTitle{\apjs}, 171, 520}

\bibitem[{{DeRosa} {et~al.}(2012){DeRosa}, {Brun}, \& {Hoeksema}}]{DeRosa2012}
{DeRosa}, M.~L., {Brun}, A.~S., \& {Hoeksema}, J.~T. 2012,
  \href{http://dx.doi.org/10.1088/0004-637X/757/1/96}{\JournalTitle{\apj}, 757,
  96}

\bibitem[{{Einfeldt}(1988)}]{Einfeldt1988}
{Einfeldt}, B. 1988, \JournalTitle{Journal of Computational Physics}, 25, 294

\bibitem[{{Evans} \& {Hawley}(1988)}]{EvansHawley1988}
{Evans}, C.~R., \& {Hawley}, J.~F. 1988,
  \href{http://dx.doi.org/10.1086/166684}{\JournalTitle{\apj}, 332, 659}

\bibitem[{{Finley} \& {Matt}(2017)}]{Finley2017}
{Finley}, A.~J., \& {Matt}, S.~P. 2017,
  \href{http://dx.doi.org/10.3847/1538-4357/aa7fb9}{\JournalTitle{\apj}, 845,
  46}

\bibitem[{{Fox} {et~al.}(2016){Fox}, {Velli}, {Bale}, {Decker}, {Driesman},
  {Howard}, {Kasper}, {Kinnison}, {Kusterer}, {Lario}, {Lockwood}, {McComas},
  {Raouafi}, \& {Szabo}}]{FoxPSP2016}
{Fox}, N.~J., {Velli}, M.~C., {Bale}, S.~D., {et~al.} 2016,
  \href{http://dx.doi.org/10.1007/s11214-015-0211-6}{\JournalTitle{\ssr}, 204,
  7}

\bibitem[{{Garraffo} {et~al.}(2016){Garraffo}, {Drake}, \&
  {Cohen}}]{Garraffo2016}
{Garraffo}, C., {Drake}, J.~J., \& {Cohen}, O. 2016,
  \href{http://dx.doi.org/10.1051/0004-6361/201628367}{\JournalTitle{\aap},
  595, A110}

\bibitem[{{Goldstein} {et~al.}(1996){Goldstein}, {Neugebauer}, {Phillips},
  {Bame}, {Gosling}, {McComas}, {Wang}, {Sheeley}, \& {Suess}}]{Goldstein1996}
{Goldstein}, B.~E., {Neugebauer}, M., {Phillips}, J.~L., {et~al.} 1996,
  \JournalTitle{\aap}, 316, 296

\bibitem[{{Hansteen} \& {Velli}(2012)}]{HansteenVelli2012}
{Hansteen}, V.~H., \& {Velli}, M. 2012,
  \href{http://dx.doi.org/10.1007/s11214-012-9887-z}{\JournalTitle{\ssr}, 172,
  89}

\bibitem[{{Holzer}(2005)}]{Holzer2005}
{Holzer}, T.~E. 2005, in ESA Special Publication, Vol. 592, Solar Wind 11/SOHO
  16, Connecting Sun and Heliosphere, ed. B.~{Fleck}, T.~H. {Zurbuchen}, \&
  H.~{Lacoste}, 115

\bibitem[{{Kopp} \& {Holzer}(1976)}]{KoppHolzer1976}
{Kopp}, R.~A., \& {Holzer}, T.~E. 1976,
  \href{http://dx.doi.org/10.1007/BF00221484}{\JournalTitle{\solphys}, 49, 43}

\bibitem[{{Lee} {et~al.}(2011){Lee}, {Luhmann}, {Hoeksema}, {Sun}, {Arge}, \&
  {de Pater}}]{Lee2011}
{Lee}, C.~O., {Luhmann}, J.~G., {Hoeksema}, J.~T., {et~al.} 2011,
  \href{http://dx.doi.org/10.1007/s11207-010-9699-9}{\JournalTitle{\solphys},
  269, 367}

\bibitem[{{Leer} {et~al.}(1982){Leer}, {Holzer}, \& {Fla}}]{Leer1982}
{Leer}, E., {Holzer}, T.~E., \& {Fla}, T. 1982,
  \href{http://dx.doi.org/10.1007/BF00213253}{\JournalTitle{\ssr}, 33, 161}

\bibitem[{{Lionello} {et~al.}(2009){Lionello}, {Linker}, \&
  {Miki{\'c}}}]{Lionello2009}
{Lionello}, R., {Linker}, J.~A., \& {Miki{\'c}}, Z. 2009,
  \href{http://dx.doi.org/10.1088/0004-637X/690/1/902}{\JournalTitle{\apj},
  690, 902}

\bibitem[{{Matsumoto} \& {Suzuki}(2012)}]{MatsumotoSuzuki2012}
{Matsumoto}, T., \& {Suzuki}, T.~K. 2012,
  \href{http://dx.doi.org/10.1088/0004-637X/749/1/8}{\JournalTitle{\apj}, 749,
  8}

\bibitem[{{McComas}(2003)}]{McComas2003}
{McComas}, D.~J. 2003, \href{http://dx.doi.org/10.1063/1.1618535}{in American
  Institute of Physics Conference Series, Vol. 679, Solar Wind Ten, ed.
  M.~{Velli}, R.~{Bruno}, F.~{Malara}, \& B.~{Bucci}}, 33

\bibitem[{{McComas} {et~al.}(2008){McComas}, {Ebert}, {Elliott}, {Goldstein},
  {Gosling}, {Schwadron}, \& {Skoug}}]{McComas2008}
{McComas}, D.~J., {Ebert}, R.~W., {Elliott}, H.~A., {et~al.} 2008,
  \href{http://dx.doi.org/10.1029/2008GL034896}{\JournalTitle{\grl}, 35,
  L18103}

\bibitem[{{McComas} {et~al.}(1998){McComas}, {Bame}, {Barraclough}, {Feldman},
  {Funsten}, {Gosling}, {Riley}, {Skoug}, {Balogh}, {Forsyth}, {Goldstein}, \&
  {Neugebauer}}]{McComas1998}
{McComas}, D.~J., {Bame}, S.~J., {Barraclough}, B.~L., {et~al.} 1998,
  \href{http://dx.doi.org/10.1029/97GL03444}{\JournalTitle{\grl}, 25, 1}

\bibitem[{{McGregor} {et~al.}(2011){McGregor}, {Hughes}, {Arge}, {Owens}, \&
  {Odstrcil}}]{McGregor2011}
{McGregor}, S.~L., {Hughes}, W.~J., {Arge}, C.~N., {Owens}, M.~J., \&
  {Odstrcil}, D. 2011,
  \href{http://dx.doi.org/10.1029/2010JA015881}{\JournalTitle{Journal of
  Geophysical Research (Space Physics)}, 116, A03101}

\bibitem[{{Mignone} {et~al.}(2007){Mignone}, {Bodo}, {Massaglia}, {Matsakos},
  {Tesileanu}, {Zanni}, \& {Ferrari}}]{Mignone2007}
{Mignone}, A., {Bodo}, G., {Massaglia}, S., {et~al.} 2007,
  \href{http://dx.doi.org/10.1086/513316}{\JournalTitle{\apjs}, 170, 228}

\bibitem[{{Miki{\'c}} {et~al.}(1999){Miki{\'c}}, {Linker}, {Schnack},
  {Lionello}, \& {Tarditi}}]{Mikic1999}
{Miki{\'c}}, Z., {Linker}, J.~A., {Schnack}, D.~D., {Lionello}, R., \&
  {Tarditi}, A. 1999,
  \href{http://dx.doi.org/10.1063/1.873474}{\JournalTitle{Physics of Plasmas},
  6, 2217}

\bibitem[{M{\"u}ller {et~al.}(2013)M{\"u}ller, Marsden, St.~Cyr, \&
  Gilbert}]{Muller2013}
M{\"u}ller, D., Marsden, R.~G., St.~Cyr, O.~C., \& Gilbert, H.~R. 2013,
  \href{http://dx.doi.org/10.1007/s11207-012-0085-7}{\JournalTitle{Solar
  Physics}, 285, 25}

\bibitem[{{Neugebauer}(1975)}]{Neugebauer1975}
{Neugebauer}, M. 1975,
  \href{http://dx.doi.org/10.1007/BF00718575}{\JournalTitle{\ssr}, 17, 221}

\bibitem[{{Parker}(1958)}]{Parker1958}
{Parker}, E.~N. 1958,
  \href{http://dx.doi.org/10.1086/146579}{\JournalTitle{\apj}, 128, 664}

\bibitem[{{Pinto} {et~al.}(2011){Pinto}, {Brun}, {Jouve}, \&
  {Grappin}}]{PintoBrun2011}
{Pinto}, R.~F., {Brun}, A.~S., {Jouve}, L., \& {Grappin}, R. 2011,
  \href{http://dx.doi.org/10.1088/0004-637X/737/2/72}{\JournalTitle{\apj}, 737,
  72}

\bibitem[{{Pinto} {et~al.}(2016){Pinto}, {Brun}, \& {Rouillard}}]{Pinto2016}
{Pinto}, R.~F., {Brun}, A.~S., \& {Rouillard}, A.~P. 2016,
  \href{http://dx.doi.org/10.1051/0004-6361/201628599}{\JournalTitle{\aap},
  592, A65}

\bibitem[{{Pinto} \& {Rouillard}(2017)}]{Pinto2017}
{Pinto}, R.~F., \& {Rouillard}, A.~P. 2017,
  \href{http://dx.doi.org/10.3847/1538-4357/aa6398}{\JournalTitle{\apj}, 838,
  89}

\bibitem[{{R{\'e}ville} {et~al.}(2015{\natexlab{a}}){R{\'e}ville}, {Brun},
  {Matt}, {Strugarek}, \& {Pinto}}]{Reville2015a}
{R{\'e}ville}, V., {Brun}, A.~S., {Matt}, S.~P., {Strugarek}, A., \& {Pinto},
  R.~F. 2015{\natexlab{a}},
  \href{http://dx.doi.org/10.1088/0004-637X/798/2/116}{\JournalTitle{\apj},
  798, 116}

\bibitem[{{R{\'e}ville} {et~al.}(2015{\natexlab{b}}){R{\'e}ville}, {Brun},
  {Strugarek}, {Matt}, {Bouvier}, {Folsom}, \& {Petit}}]{Reville2015b}
{R{\'e}ville}, V., {Brun}, A.~S., {Strugarek}, A., {et~al.} 2015{\natexlab{b}},
  \href{http://dx.doi.org/10.1088/0004-637X/814/2/99}{\JournalTitle{\apj}, 814,
  99}

\bibitem[{{R{\'e}ville} {et~al.}(2016{\natexlab{a}}){R{\'e}ville}, {Folsom},
  {Strugarek}, \& {Brun}}]{Reville2016}
{R{\'e}ville}, V., {Folsom}, C.~P., {Strugarek}, A., \& {Brun}, A.~S.
  2016{\natexlab{a}},
  \href{http://dx.doi.org/10.3847/0004-637X/832/2/145}{\JournalTitle{\apj},
  832, 145}

\bibitem[{{R{\'e}ville} {et~al.}(2016{\natexlab{b}}){R{\'e}ville}, {Folsom},
  {Strugarek}, \& {Brun}}]{Reville2016csss}
{R{\'e}ville}, V., {Folsom}, C.~P., {Strugarek}, A., \& {Brun}, A.~S.
  2016{\natexlab{b}}, \href{http://dx.doi.org/10.5281/zenodo.160714}{in 19th
  Cambridge Workshop on Cool Stars, Stellar Systems, and the Sun (CS19)}, 33

\bibitem[{{Riley} {et~al.}(2006){Riley}, {Linker}, {Miki{\'c}}, {Lionello},
  {Ledvina}, \& {Luhmann}}]{Riley2006}
{Riley}, P., {Linker}, J.~A., {Miki{\'c}}, Z., {et~al.} 2006,
  \href{http://dx.doi.org/10.1086/508565}{\JournalTitle{\apj}, 653, 1510}

\bibitem[{{Schatten} {et~al.}(1969){Schatten}, {Wilcox}, \&
  {Ness}}]{Schatten1969}
{Schatten}, K.~H., {Wilcox}, J.~M., \& {Ness}, N.~F. 1969,
  \href{http://dx.doi.org/10.1007/BF00146478}{\JournalTitle{\solphys}, 6, 442}

\bibitem[{{Scherrer} {et~al.}(1977){Scherrer}, {Wilcox}, {Svalgaard}, {Duvall},
  {Dittmer}, \& {Gustafson}}]{Scherrer1977}
{Scherrer}, P.~H., {Wilcox}, J.~M., {Svalgaard}, L., {et~al.} 1977,
  \href{http://dx.doi.org/10.1007/BF00159925}{\JournalTitle{\solphys}, 54, 353}

\bibitem[{{Schrijver} \& {De Rosa}(2003)}]{SchrijverDeRosa2003}
{Schrijver}, C.~J., \& {De Rosa}, M.~L. 2003,
  \href{http://dx.doi.org/10.1023/A:1022908504100}{\JournalTitle{\solphys},
  212, 165}

\bibitem[{{Schwenn}(1983)}]{Schwenn1983}
{Schwenn}, R. 1983, in NASA Conference Publication, Vol. 228, NASA Conference
  Publication

\bibitem[{{Smith}(2011)}]{Smith2011}
{Smith}, E.~J. 2011,
  \href{http://dx.doi.org/10.1016/j.jastp.2010.03.019}{\JournalTitle{Journal of
  Atmospheric and Solar-Terrestrial Physics}, 73, 277}

\bibitem[{{Smith} \& {Balogh}(1995)}]{SmithBalogh1995}
{Smith}, E.~J., \& {Balogh}, A. 1995,
  \href{http://dx.doi.org/10.1029/95GL02826}{\JournalTitle{\grl}, 22, 3317}

\bibitem[{{Sokolov} {et~al.}(2013){Sokolov}, {van der Holst}, {Oran}, {Downs},
  {Roussev}, {Jin}, {Manchester}, {Evans}, \& {Gombosi}}]{Sokolov2013}
{Sokolov}, I.~V., {van der Holst}, B., {Oran}, R., {et~al.} 2013,
  \href{http://dx.doi.org/10.1088/0004-637X/764/1/23}{\JournalTitle{\apj}, 764,
  23}

\bibitem[{{Suzuki}(2006)}]{Suzuki2006}
{Suzuki}, T.~K. 2006,
  \href{http://dx.doi.org/10.1086/503102}{\JournalTitle{\apjl}, 640, L75}

\bibitem[{{Suzuki} \& {Inutsuka}(2006)}]{SuzukiInutsuka2006}
{Suzuki}, T.~K., \& {Inutsuka}, S.-I. 2006,
  \href{http://dx.doi.org/10.1029/2005JA011502}{\JournalTitle{Journal of
  Geophysical Research (Space Physics)}, 111, 6101}

\bibitem[{{Svalgaard} {et~al.}(1978){Svalgaard}, {Duvall}, \&
  {Scherrer}}]{Svalgaard1978}
{Svalgaard}, L., {Duvall}, Jr., T.~L., \& {Scherrer}, P.~H. 1978,
  \href{http://dx.doi.org/10.1007/BF00157268}{\JournalTitle{\solphys}, 58, 225}

\bibitem[{{Usmanov}(1993)}]{Usmanov1993}
{Usmanov}, A.~V. 1993,
  \href{http://dx.doi.org/10.1007/BF00662021}{\JournalTitle{\solphys}, 146,
  377}

\bibitem[{{Usmanov} {et~al.}(2000){Usmanov}, {Goldstein}, {Besser}, \&
  {Fritzer}}]{Usmanov2000}
{Usmanov}, A.~V., {Goldstein}, M.~L., {Besser}, B.~P., \& {Fritzer}, J.~M.
  2000, \href{http://dx.doi.org/10.1029/1999JA000233}{\JournalTitle{\jgr}, 105,
  12675}

\bibitem[{{van der Holst} {et~al.}(2014){van der Holst}, {Sokolov}, {Meng},
  {Jin}, {Manchester}, {T{\'o}th}, \& {Gombosi}}]{vanDerHolst2014}
{van der Holst}, B., {Sokolov}, I.~V., {Meng}, X., {et~al.} 2014,
  \href{http://dx.doi.org/10.1088/0004-637X/782/2/81}{\JournalTitle{\apj}, 782,
  81}

\bibitem[{{Velli}(2010)}]{Velli2010}
{Velli}, M. 2010,
  \href{http://dx.doi.org/10.1063/1.3395823}{\JournalTitle{Twelfth
  International Solar Wind Conference}, 1216, 14}

\bibitem[{{Wang}(1998)}]{Wang1998}
{Wang}, Y.-M. 1998, in Astronomical Society of the Pacific Conference Series,
  Vol. 154, Cool Stars, Stellar Systems, and the Sun, ed. R.~A. {Donahue} \&
  J.~A. {Bookbinder}, 131

\bibitem[{{Wang}(2010)}]{Wang2010}
{Wang}, Y.-M. 2010,
  \href{http://dx.doi.org/10.1088/2041-8205/715/2/L121}{\JournalTitle{\apjl},
  715, L121}

\bibitem[{{Wang}(2016)}]{Wang2016}
---. 2016,
  \href{http://dx.doi.org/10.3847/1538-4357/833/1/121}{\JournalTitle{\apj},
  833, 121}

\bibitem[{{Wang} {et~al.}(2009){Wang}, {Ko}, \& {Grappin}}]{Wang2009}
{Wang}, Y.-M., {Ko}, Y.-K., \& {Grappin}, R. 2009,
  \href{http://dx.doi.org/10.1088/0004-637X/691/1/760}{\JournalTitle{\apj},
  691, 760}

\bibitem[{{Wang} \& {Sheeley}(1990)}]{WangSheeley1990}
{Wang}, Y.-M., \& {Sheeley}, Jr., N.~R. 1990,
  \href{http://dx.doi.org/10.1086/168805}{\JournalTitle{\apj}, 355, 726}

\bibitem[{{Wang} \& {Sheeley}(1991)}]{WangSheeley1991}
---. 1991, \href{http://dx.doi.org/10.1086/186020}{\JournalTitle{\apjl}, 372,
  L45}

\bibitem[{{Withbroe}(1988)}]{Withbroe1988}
{Withbroe}, G.~L. 1988,
  \href{http://dx.doi.org/10.1086/166015}{\JournalTitle{\apj}, 325, 442}

\end{thebibliography}
\end{document}